\documentclass[twocolumn]{aastex6}
\usepackage{graphicx}
\usepackage[countmax]{subfloat}

\begin{document}
\title{X-ray Intraday Variability of Five TeV Blazars with \textit{NuSTAR}}
\author{Ashwani Pandey\altaffilmark{1,2}, Alok C. Gupta\altaffilmark{3,1}, \& Paul J. Wiita\altaffilmark{4}}

\altaffiltext{1}{Aryabhatta Research Institute of Observational Sciences (ARIES), Manora Peak, Nainital 263002, India; ashwanitapan@gmail.com}
\altaffiltext{2}{Department of Physics, DDU Gorakhpur University, Gorakhpur 273009, India}
\altaffiltext{3}{Key Laboratory for Research in Galaxies and Cosmology, Shanghai Astronomical Observatory, Chinese Academy of Sciences, 80 Nandan Road, Shanghai 200030, China; acgupta30@gmail.com}
\altaffiltext{4}{Department of Physics, The College of New Jersey, 2000 Pennington Rd., Ewing, NJ 08628-0718, USA; wiitap@tcnj.edu}

\begin{abstract}

We have examined 40 \textit{NuSTAR} light curves (LCs) of five TeV emitting high synchrotron peaked 
blazars: 1ES 0229$+$200, Mrk 421, Mrk 501, 1ES 1959$+$650 and PKS 2155$-$304.  Four of the blazars
showed intraday variability in the \textit{NuSTAR} energy range of 3--79 keV. 
Using an auto correlation function analysis we searched for intraday variability timescales in these LCs and found
indications of several between 2.5 and 32.8 ks in eight LCs of Mrk 421, a timescale around 8.0 ks for one LC of Mrk 501, 
and timescales of 29.6 ks and 57.4 ks in two LCs of PKS 2155-304.
The other two blazars' LCs do not show any evidence for intraday variability timescales shorter than the lengths 
of those observations; however, the data was both sparser and noisier, for  them.
We found positive correlations with zero lag between soft (3--10 keV) and hard (10--79 keV) bands for most of the LCs, 
indicating that their emissions originate from the same electron population. We examined spectral variability using a hardness 
ratio analysis and noticed a general ``harder-when-brighter" behavior.
The 22  LCs of Mrk 421 observed between July 2012 and April 2013 show that this source was in a quiescent 
state for an extended period of 
time and then underwent an unprecedented double peaked outburst while monitored on a daily basis during 
 10 -- 16 April 2013.  We briefly discuss models capable of explaining  these blazar emissions.

\end{abstract}
\keywords {BL Lacertae objects: general -- BL Lacertae objects: individual (1ES 0229$+$200, Mrk 421, Mrk 501, 
1ES 1959$+$650, PKS 2155$-$304)}

\section{Introduction} \label{sec:intro}
Blazars are the subclass of active galactic nuclei (AGN) understood to have relativistic jets pointing close 
($\leq 10^{\circ}$; \cite{1995PASP..107..803U}) to line of sight of the observer and their spectra are dominated 
by non-thermal radiation coming from the jets. Blazars are classically classified on the basis of the presence or 
absence of emission lines in their optical spectra: BL Lacertae (BL Lac) objects have no,  or very weak, emission 
lines (EW $< 5\AA$) (\cite{1991ApJS...76..813S}; \cite{1996MNRAS.281..425M}), while flat spectrum radio quasars 
(FSRQs) have strong broad emission lines. The spectral energy distributions (SEDs) of blazars have two broad bumps, 
which leads to another classification as low synchrotron peaked (LSP), intermediate synchrotron peaked (ISP), or 
high synchrotron peaked (HSP) blazars.  The first (low energy) peak lies in the infrared to optical region in LSPs 
(comprised of FSRQs and low-frequency peaked BL Lac objects, or LBLs) and in the far ultraviolet to X-ray energies 
in HSPs (high-frequency peaked BL Lac objects, or HBLs). The second (high energy) bump is in the gamma-ray band, 
peaking at GeV energies in LSPs and at TeV energies in HSPs. \cite{2010ApJ...716...30A} classified LSPs, ISPs and 
HSPs on the basis of their synchrotron peak frequency, $\nu_s$, as  having $\nu_s \leq 10^{14}$Hz, 
1$0^{14}$ Hz $< \nu_s < 10^{15}$Hz, and $\nu_s \geq 10^{15}$Hz, respectively. 

\par It is well accepted that the low-energy peak is due to synchrotron emission from ultra-relativistic electrons 
in the jet. The physical origin of the high-energy peak is still under some debate and both leptonic and hadronic 
models have been proposed to explain it.   In the generally favored leptonic models the high-energy peak is 
attributed to the inverse-Compton scattering of either synchrotron or ambient photons by the same relativistic 
electrons responsible for the synchrotron emission, while hadronic models attribute the gamma-ray peak to the direct 
proton and muon synchrotron radiations (e.g. \cite{2007Ap&SS.307...69B}).  

\par Blazars emit radiations at all observable wavelengths from radio to very high energy (VHE) gamma-rays and 
exhibit rapid and strong flux variability on different timescales ranging from few tens of seconds to years. Blazar 
variability timescales are often classified as: intra-day variability (IDV), also called microvariability or 
intra-night variability for detectable changes seen to occur over less than a day \citep{1995ARA&A..33..163W});  short term 
variability (STV) for fluctuations over a few days to a few months; and long term variability (LTV) for changes 
seen over  longer periods  \citep{2004A&A...422..505G}).

The Nuclear Spectroscopic Telescope Array (\textit{NuSTAR}) is the first X-ray telescope that observes in throughout the critical 
energy range of 3--79 keV \citep{2013ApJ...770..103H}. It has two co-aligned hard X-ray grazing incidence telescopes and 
two independent solid state focal plane detectors, referred to as the Focal Plane Module A (FPMA) and Focal Plane Module 
B (FPMB), which were designed to have identical detection efficiencies.  \textit{NuSTAR} has an angular resolution (FWHM) of 
18 arcsec and a good energy resolution (FWHM) of 400eV at 10 keV and 
900 eV at 60 keV. There is no photon pile-up problem as these detectors do not make use of integrating CCD readouts;
 hence, it is particularly good  for timing analyses. Thanks to its unprecedented sensitivity in the hard X-ray band,  \textit{NuSTAR} can be particularly helpful
in providing understanding of the  high 
energy radiation mechanism, as this band is near where the first and second bumps in the SED intersect in HSPs.

 Until 2005, only 6 TeV HBLs (Mrk 421, Mrk 501, 1ES 1426+428, 1ES 1959+650, PKS 2155$-$304, and 1ES 2344+514) 
were known. Thanks to the discovery of HBLs over the last decade by the \textit{Fermi} satellite and several ground-based
very high energy (VHE) $\gamma-$ray 
facilities (e.g. {\it HESS (High Energy Stereoscopic System), MAGIC (Major Atmospheric Gamma-ray Imaging Cherenkov 
Telescopes), VERITAS (Very Energetic Radiation Imaging Telescope Array System)}, etc.), $\gamma-$ray blazar 
astronomy has undergone a revolution. 

\cite{2005A&A...440..855G} performed a statistical analysis of the optical IDV for various classes 
of AGNs and reported that LBL blazars show IDV during $\sim$ 60--65\% of nights if observed continuously for less than 
6 hours but this fraction rises to $\sim$ 80--85\%  if observed for more than 6 hours. In a pilot search for optical IDV of HBLs, 
\citep{{2012AJ....143...23G},{2012MNRAS.425.3002G},{2012MNRAS.420.3147G}}, 
found that of the 144 LCs they measured, only 6 ($\sim$ 4\%) showed IDV, a significantly smaller fraction than for LBLs.
A  study of the soft X-ray IDV  of LBLs \citep{{2015MNRAS.451.1356K},{2016MNRAS.462.1508G}} 
  found that out of 50 IDV LCs only 2  ($\sim$ 4\%) showed IDV.   On the other hand, HBLs are highly variable 
in soft X-rays, e.g. \citep{2010ApJ...718..279G}, so these classes of blazars exhibit very different behaviors in the optical and X-ray bands. 

The most puzzling aspect of the flux variability of blazars is on IDV timescales and an examination of these changes
in the unexplored hard X-ray band  provides the main motivation  for this work. We focus on those blazars that show
emission at the highest energies, the HBL TeV blazars.
There are a  total of 61 TeV blazars\footnote{\url{http://tevcat.uchicago.edu/}} known at the time of writing, of which 47
are HBLs, 8 are IBLs, 1 is an LBL and 5 are FSRQs. Of the 47 TeV HBLs, \textit{NuSTAR} had observed only five of them
for a total of 43 pointed observations. We reduced all the available archival data for these
objects to produce LCs and  then used auto correlation function (ACF) analyses  to search for  possible timescales of 
variability in the \textit{NuSTAR} LCs of these five TeV HBLs.  The extensive data taken on Mrk 421 gave us the additional opportunity to look at its STV. 

\par The outline of the paper is as follows. We briefly describe the data selection criteria and the data reduction 
method in section \ref{sec:data} and discuss the techniques used to search for flux and spectral variability properties in section 
\ref{sec:analysis}. In section \ref{sec:result}, the results are given. A discussion and our conclusions are given in sections
\ref{sec:discussion} and \ref{conclusion}, respectively.

\section{\textit{NuSTAR} DATA SELECTION AND PROCESSING} \label{sec:data}

\subsection{Data Selection Criteria} \label{subsec:dataselect}

We have examined the emission of all of the five TeV HBLs observed by  \textit{NuSTAR}:  1ES 0229$+$200, Markarian (Mrk) 421, 
Mrk 501, 1ES 1959$+$650 and PKS 2155$-$304.  
This gave us a new opportunity to study the variability nature of 
these blazars in hard X-ray energies.   In order to search for IDV we only selected observations for which the good exposure times
exceeded 5 ks.  We downloaded all 43 data sets publicly available from the HEASARC Data 
Archive\footnote{\url{http://heasarc.gc.nasa.gov/docs/archive.html}}, and
 our minimum temporal constraint yielded 3 observations for 1ES 0229$+$200, 22 for Mrk 421, 
4 for Mrk 501, 2 for 1ES 1959$+$650 and 9 for PKS 2155$-$304. These 40 observations were made between 7 July 2012 and 
22 September 2014 and the good exposure times ranged  from 5.76 ks to 57.51 ks. The observing log of \textit{NuSTAR} data for these five TeV blazars 
is given in Table \ref{tab:obs_log}.

\begin{table*}
\centering
\caption{Observation log of \textit{NuSTAR} data for five TeV HBLs.}
 \label{tab:obs_log}
 \begin{tabular}{lcclcc}
  \hline
Blazar Name 	   & ~~Obs. Date  & Start Time (UT) & ~~~Obs. ID 	      & Total Elapsed & Good Exposure \\
                   & ~~yyyy-mm-dd & hh:mm:ss    &                &      Time (ks)    &  Time (ks)       \\
\hline
1ES 0229+200	   &  2013-10-02 & 00:06:07   &  60002047002   &  ~32.43   &   16.26  \\
		   &  2013-10-05 & 23:31:07   &  60002047004   &  ~38.24   &   20.29  \\
		   &  2013-10-10 & 23:11:07   &  60002047006   &  ~32.35   &   18.02  \\

Mrk 421		   &  2012-07-07 & 01:56:07   &  10002015001   &   ~78.45   &   42.03  \\
		   &  2012-07-08 & 01:46:07   &  10002016001   &   ~44.16   &   24.89  \\
		   &  2013-01-02 & 18:41:07   &  60002023002   &   ~14.51   &   ~9.15  \\
		   &  2013-01-10 & 01:16:07   &  60002023004   &   ~43.44   &   22.63  \\
		   &  2013-01-15 & 00:56:07   &  60002023006   &   ~44.14   &   24.18  \\
		   &  2013-01-20 & 02:21:07   &  60002023008   &   ~44.14   &   24.97  \\
		   &  2013-02-06 & 00:16:07   &  60002023010   &   ~42.09   &   19.31  \\
		   &  2013-02-12 & 00:16:07   &  60002023012   &   ~35.39   &   14.78  \\
		   &  2013-02-16 & 23:36:07   &  60002023014   &   ~37.14   &   17.36  \\
		   &  2013-03-04 & 23:06:07   &  60002023016   &   ~34.99   &   17.25  \\
		   &  2013-03-11 & 23:01:07   &  60002023018   &   ~31.88   &   17.47  \\
		   &  2013-03-17 & 00:11:07   &  60002023020   &   ~35.08   &   16.56  \\
		   &  2013-04-02 & 17:16:07   &  60002023022   &   ~54.19   &   24.77  \\
		   &  2013-04-10 & 21:26:07   &  60002023024   &   ~12.88   &   ~5.76  \\
		   &  2013-04-11 & 01:01:07   &  60002023025   &  117.29   &   57.51  \\
		   &  2013-04-12 & 20:36:07   &  60002023027   &   ~18.69   &   ~7.63  \\
		   &  2013-04-13 & 21:36:07   &  60002023029   &   ~32.57   &   16.51  \\
		   &  2013-04-14 & 21:41:07   &  60002023031   &   ~32.58   &   15.61  \\
		   &  2013-04-15 & 22:01:07   &  60002023033   &   ~32.56   &   17.28  \\
		   &  2013-04-16 & 22:21:07   &  60002023035   &   ~38.11   &   20.28  \\
		   &  2013-04-18 & 00:16:07   &  60002023037   &   ~31.09   &   17.80  \\
		   &  2013-04-19 & 00:31:07   &  60002023039   &   ~26.74   &   15.96  \\

Mrk 501		   &  2013-04-13 & 02:31:07   &  60002024002   &   ~35.76   &   18.28  \\
		   &  2013-05-08 & 20:01:07   &  60002024004   &   ~55.20   &   26.14  \\
		   &  2013-07-12 & 21:31:07   &  60002024006   &   ~20.90   &   10.86  \\
		   &  2013-07-13 & 20:16:07   &  60002024008   &   ~20.71   &   10.34  \\

1ES 1959$+$650	   &  2014-09-17 & 02:36:07   &  60002055002   &   ~35.04   &   19.61  \\
		   &  2014-09-22 & 02:06:07   &  60002055004   &   ~32.80   &   20.34  \\

PKS 2155$-$304	   &  2012-07-08 & 14:36:07   &  10002010001   &   ~71.57   &   33.84  \\
		   &  2013-04-23 & 19:46:07   &  60002022002   &   ~90.11   &   45.06  \\
		   &  2013-07-16 & 22:51:07   &  60002022004   &   ~26.21   &   13.86  \\
		   &  2013-08-02 & 21:51:07   &  60002022006   &   ~29.93   &   10.97  \\
		   &  2013-08-08 & 22:01:07   &  60002022008   &   ~36.54   &   13.50  \\
		   &  2013-08-14 & 21:51:07   &  60002022010   &   ~31.51   &   10.53  \\
		   &  2013-08-26 & 19:51:07   &  60002022012   &   ~24.33   &   11.36  \\
		   &  2013-09-04 & 21:56:07   &  60002022014   &   ~30.49   &   12.28  \\
		   &  2013-09-28 & 22:56:07   &  60002022016   &   ~25.66   &   11.53  \\

\hline

\end{tabular}
\end{table*}

\subsection{Data Reduction}\label{subsec:datareduce}

We reduced and analyzed the \textit{NuSTAR} data using HEASOFT\footnote{\url{http://heasarc.gc.nasa.gov/docs/nustar/analysis/}}
 version 6.17 and CALDB version 20151008. 
We used the standard \textit{nupipeline} script to generate calibrated, cleaned and screened level 2 event 
files. Each source LC is then extracted, using the \textit{nuproducts} script, from a circular 
region centered at the source.  We employ a circular background region that is selected to be both 
relatively close to the source but also  far enough away to be free from contamination by the source. The radii of the source 
and background regions for our five TeV HBLs are listed in Table \ref{tab:region}. 
Since \textit{NuSTAR} has two co-aligned and nearly identical detectors, FPMA and FPMB, their 
count rates  are background-subtracted and summed to generate the final light curves. 
We used a bin size of 5 minutes to extract the finer binned light curves.   This sampling interval is similar
to those employed in most ground-based optical IDV studies of AGN.
 
\begin{table}
\caption{Source and background region sizes}
 \label{tab:region}
 \begin{tabular}{lcc}
  \hline
Blazar Name & Source radius  & Background radius  \\
\hline
1ES 0229$+$200 & $20^{\prime\prime}$  & $30^{\prime\prime}$  \\
Mrk 421 & $30^{\prime\prime}$   & $70^{\prime\prime}$  \\
Mrk 501 & $40^{\prime\prime}$   & $40^{\prime\prime}$  \\
1ES 1959$+$650  & $30^{\prime\prime}$  & $30^{\prime\prime}$  \\
PKS 2155$-$304  & $30^{\prime\prime}$  & $30^{\prime\prime}$  \\
\hline
\end{tabular}
\end{table}

\begin{subfigures} 
\label{fig:fig1}
\begin{figure*}
\centering

\includegraphics[width=18cm, height=20cm]{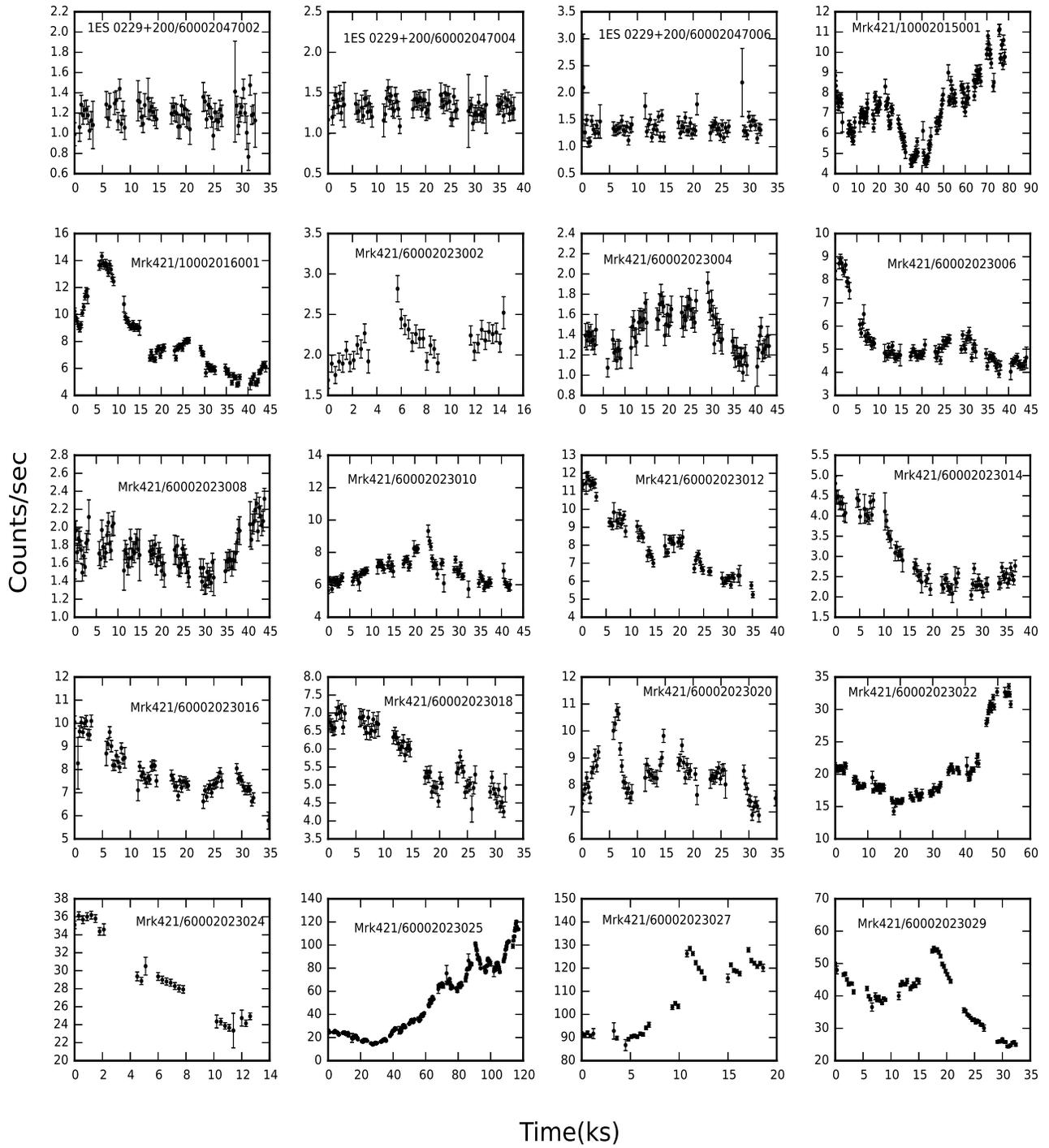}
\caption{\label{1_a}\textit{NuSTAR} light curves of the TeV HBLs 1ES 0229$+$200 and Mrk 421.  The name of the HBL and the observation ID are given in each plot.} 
\end{figure*}

\begin{figure*}
\centering
\includegraphics[width=18cm, height=20cm]{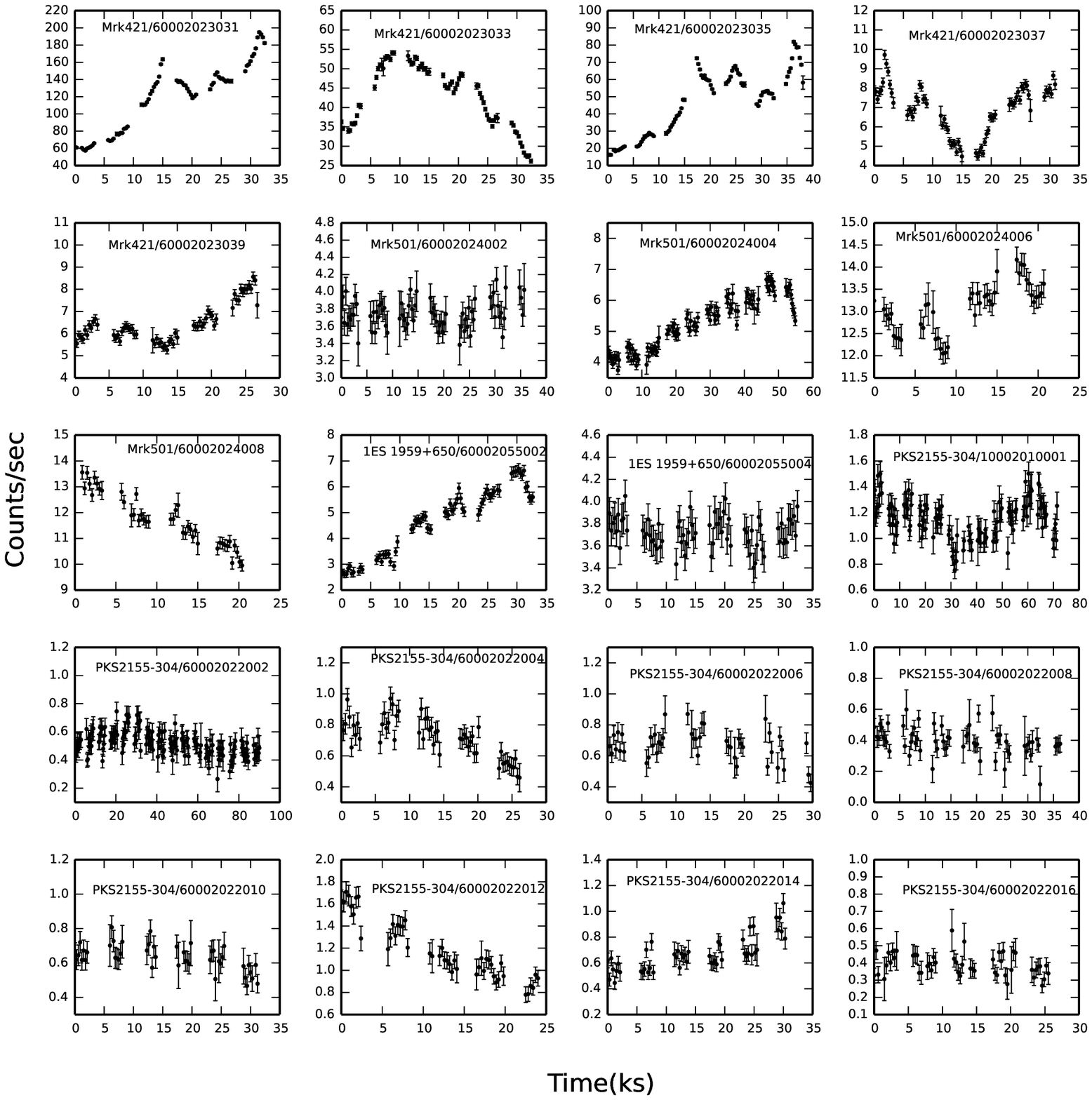}
 \caption{\label{1_b}\textit{NuSTAR} light curves of the TeV HBLs Mrk 421, Mrk 501, 1ES 1959$+$650 and PKS 2155$-$304. The name of the HBL  and the observation ID are given in each plot.}
\end{figure*}
\end{subfigures}


\section{ANALYSIS TECHNIQUES} \label{sec:analysis}
\subsection{Excess Variance} \label{subsec:frac}
Blazars are known to exhibit rapid and strong flux variations on diverse timescales across the complete electromagnetic 
spectrum. The strength of this variability is often quantified by calculating the excess variance, which is a measure 
of a source's intrinsic variance.
The excess variance is calculated by subtracting the variance arising from measurement errors from the total variance of the observed
LC. If a LC consisting of $N$ measured flux values, $x_i$, contains corresponding finite uncertainties $\sigma_{err,i}$ arising from 
measurement errors, then the excess variance is calculated (e.g., \cite{2003MNRAS.345.1271V}) as 
\begin{equation}
\sigma_{XS}^2 = S^2 - \overline{\sigma_{err}^2},
\end{equation}
where $\overline{\sigma_{err}^2}$ is the mean square error, given by, 
\begin{equation} 
\overline{\sigma_{err}^2} =\frac{1}{N} \sum\limits_{i=1}^N \sigma^2_{err,i} .
\end{equation}
The quantity $S^2$ is the sample variance of the LC, and is given by
\begin{equation}
S^2 = \frac{1}{N-1} \sum\limits_{i=1}^N (x_i - \bar{x})^2 ,
\end{equation}
where $\bar{x}$ is the arithmetic mean of $x_i$. \\
The normalized excess variance is $ \sigma^2_{NXS} = {\sigma^2_{XS}} / {\bar{x}^2}$ 
and the fractional rms variability amplitude, $F_{var}$, which is the square root of $\sigma^2_{NXS}$, is thus
\begin{equation}
F_{var} = \sqrt{\frac{S^2 - \overline{\sigma_{err}^2}}{{\bar{x}^2}}} .
\end{equation}
The uncertainty on $F_{var}$ is given by (e.g., \cite{2003MNRAS.345.1271V}) 
\begin{equation}
err(F_{var}) =  \sqrt{\left( \sqrt{\frac{1}{2N}}\frac{\overline{\sigma_{err}^2}}{\bar{x}^2 F_{var}} \right)  ^ 2+ \left(  \sqrt{\frac{\overline{\sigma_{err}^2}}{N}} \frac{1}{\bar{x}}\right) ^2 }.
\end{equation}

\subsection{Discrete Correlation Functions}
\label{sec:DCF}

The advantage of using a Discrete Correlation Function (DCF) over a classical correlation function is that it can be applied to unevenly sampled data sets, as are typical of astronomical observations and as we have here, without interpolating  between data \citep{1988ApJ...333..646E}. The DCF  is  calculated in the following fashion.
First, we calculate the set of unbinned discrete correlations for two discrete data sets $x$ and  $y$ (e.g. \cite{2015A&A...582A.103G}) as
\begin{equation}
 UDCF_{ij}=\frac{(x_i - \bar{x})(y_j - \bar{y})}{\sqrt{\sigma_x^2 \sigma_y^2}},
\end{equation}
where $x_i$ and $y_j$ are the data points, $\bar{x}$ and $\bar{y}$ are their means, and $\sigma_x$  and  $\sigma_y$  are their standard deviations, respectively. Each of these is associated with the pairwise lag $\Delta t_{ij} = t_j - t_i$.
After calculating the UDCF, the correlation function is binned in time. The method does not automatically define a bin size so one must investigate several values for this parameter. If the bin size is too large, information is lost but if the bin size is too small, we can get spurious correlations, and the time scales may be difficult to interpret. Simulations conducted by \cite{1988ApJ...333..646E} suggest that the results depend only weakly on the choice of bin size.

\par After binning, the DCF can be calculated by averaging the UDCF values ($M$ in number) for which $\tau - \Delta\tau/2  \leq  \Delta t_{ij} < \tau + \Delta \tau/2$ as,

\begin{equation}
DCF(\tau)=\frac{1}{M}\sum  UDCF_{ij} .
\end{equation}
\cite{1988ApJ...333..646E} defined the standard error for each bin  as 
\begin{equation}
\sigma_{DCF}(\tau) = \frac{\sqrt{\sum[UDCF_{ij} - DCF(\tau)]^2}}{M - 1} .
\end{equation}

In general, a positive DCF peak means that the two data signals are correlated, while
a negative DCF peak means that he two data sets are anti-correlated, but
no DCF peak, or DCF = 0, means that no correlation exists between the two data sets.
When correlating a data series with itself (i.e., $x = y$), we obtain the auto-correlation function (ACF) with an automatic peak at $\tau = 0$, indicating the absence of any time lag.  For an ACF any other strong peak can indicate the presence of periodicity, but strong dips provide an indication of the presence and value of a timescale in the data \citep{2011MNRAS.413.2157R}.

\subsection{Hardness Ratio}
\label{sec:HR}
To examine the spectral variability of the {\it NuSTAR} X-ray emission from the five TeV HBLs, we split the LCs into two energy bands: a 
soft  band from 3--10 keV and a hard  band from 10-79 keV.  We  then computed the hardness ratio (HR) as 
\begin{equation}
HR = \frac{(H-S)}{(H+S)},
\end{equation} 
where H and S are the net count rates in the hard (10--79 keV) and soft (3--10 keV) bands, respectively.  The hardness ratio is a commonly used and simple model-independent method to study spectral variations. We examined variations of the HRs with time to search for  spectral changes over this broad X-ray band.


\begin{subfigures}
\label{fig:fig2}

\begin{figure*}
\centering
\includegraphics[width=18cm, height=20cm]{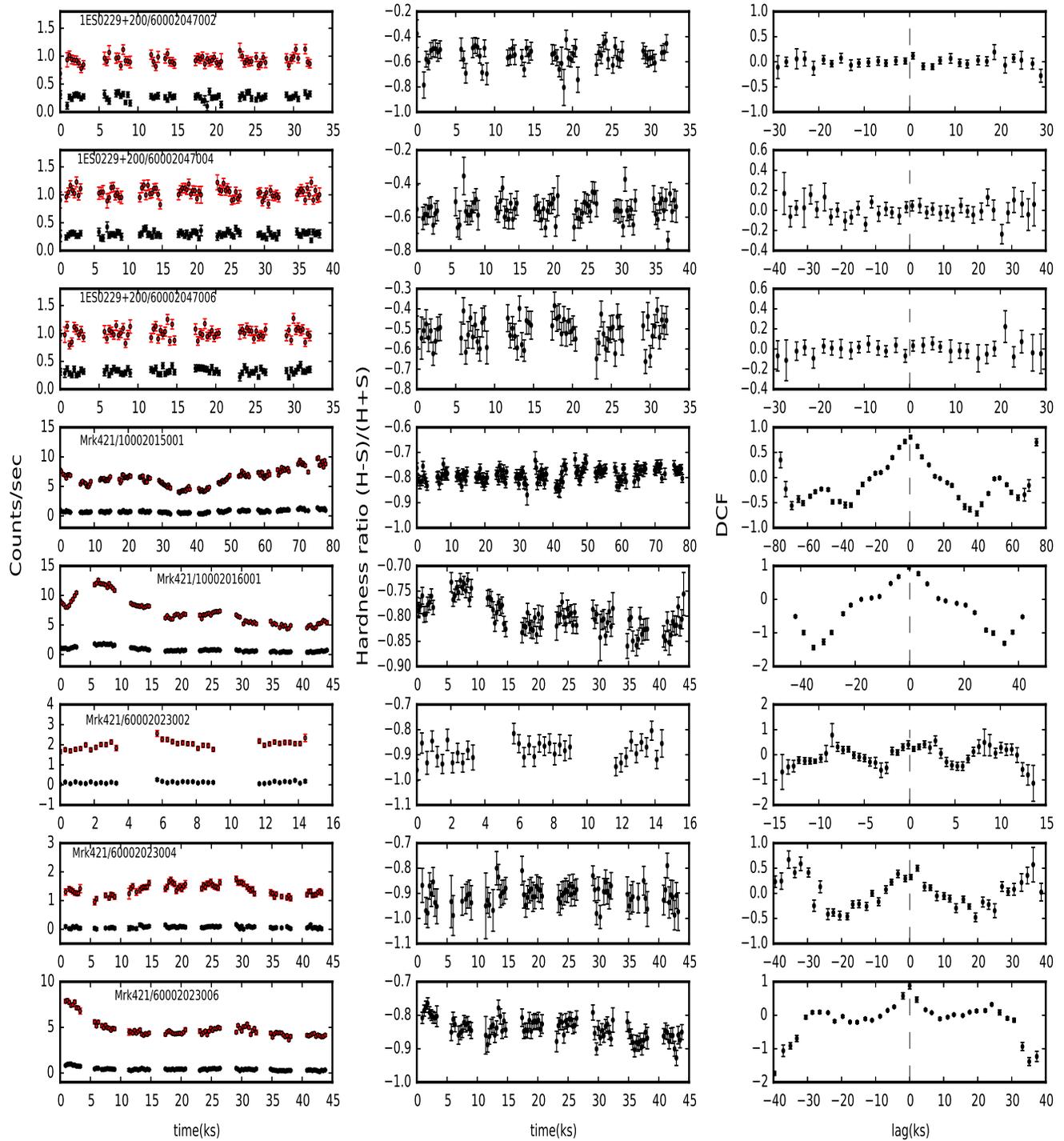}
\caption{\label{2_a} Soft (3--10 keV, denoted by red filled circles) and hard (10--79 keV, denoted by black filled circles) LCs (left panels), hardness ratios (middle panels), and the discrete correlation functions between soft and hard LCs (right panels) of the blazars 1ES 0229$+$200 and Mrk 421. The source names and observation ids are given in the left panels.}
\end{figure*}

\begin{figure*}
\centering
\includegraphics[width=18cm, height=20cm]{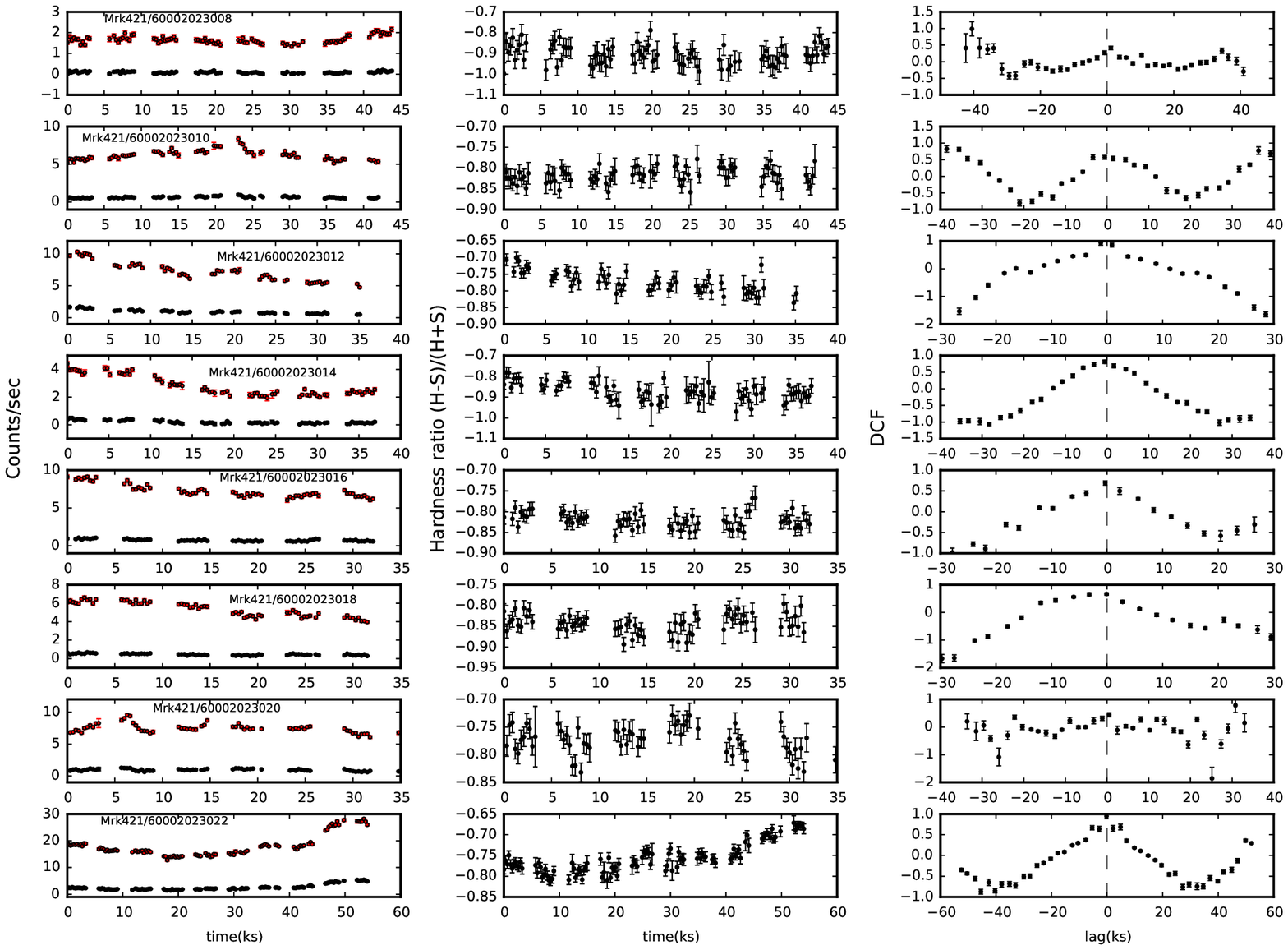}
\caption{\label{2_b} As in Figure \ref{2_a} for Mrk 421.}
\end{figure*}

\begin{figure*}
\centering
\includegraphics[width=18cm, height=20cm]{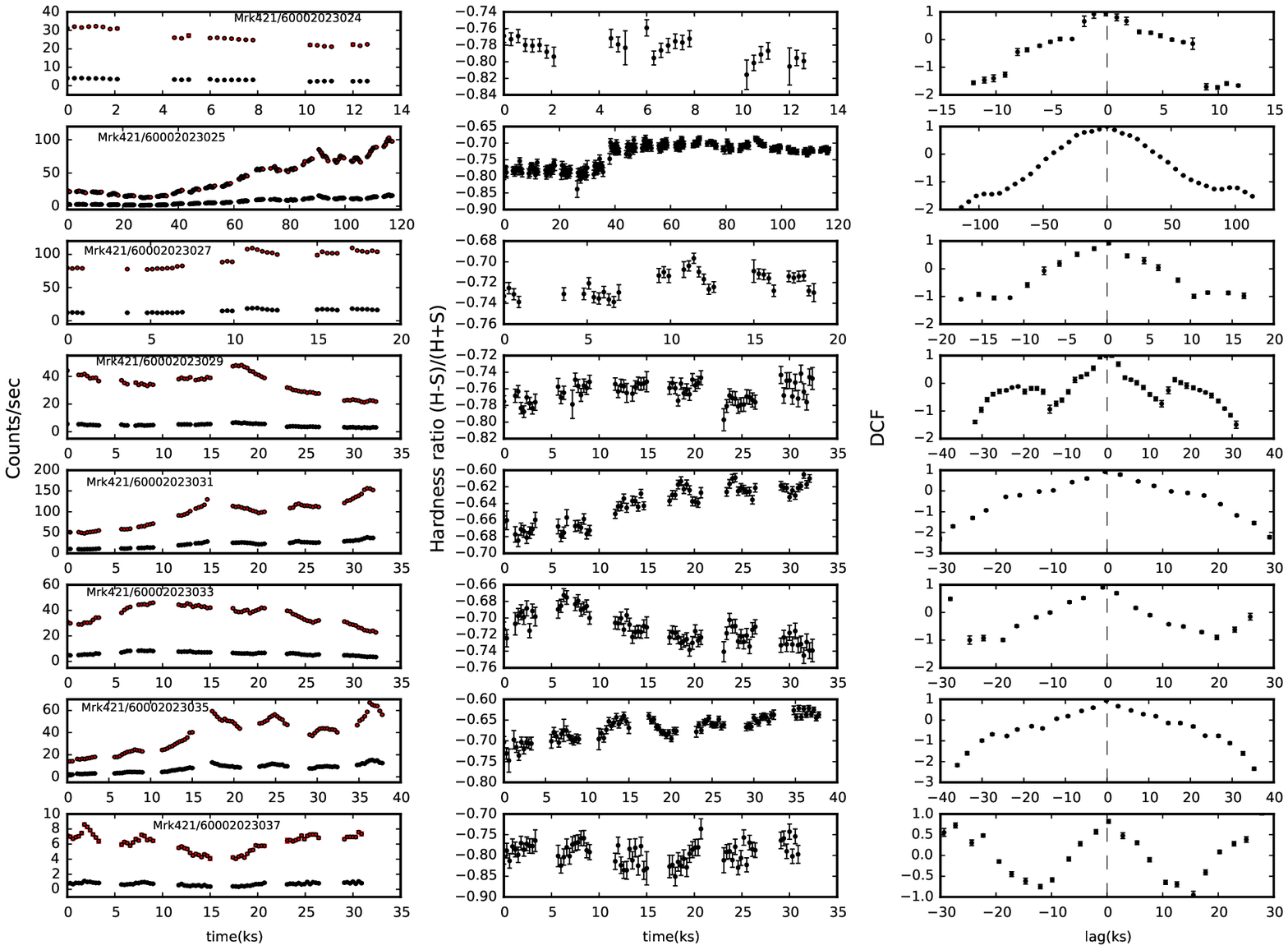}
\caption{\label{2_c} As in Figure \ref{2_a} for Mrk 421.}
\end{figure*}

\begin{figure*}
\centering
\includegraphics[width=18cm, height=20cm]{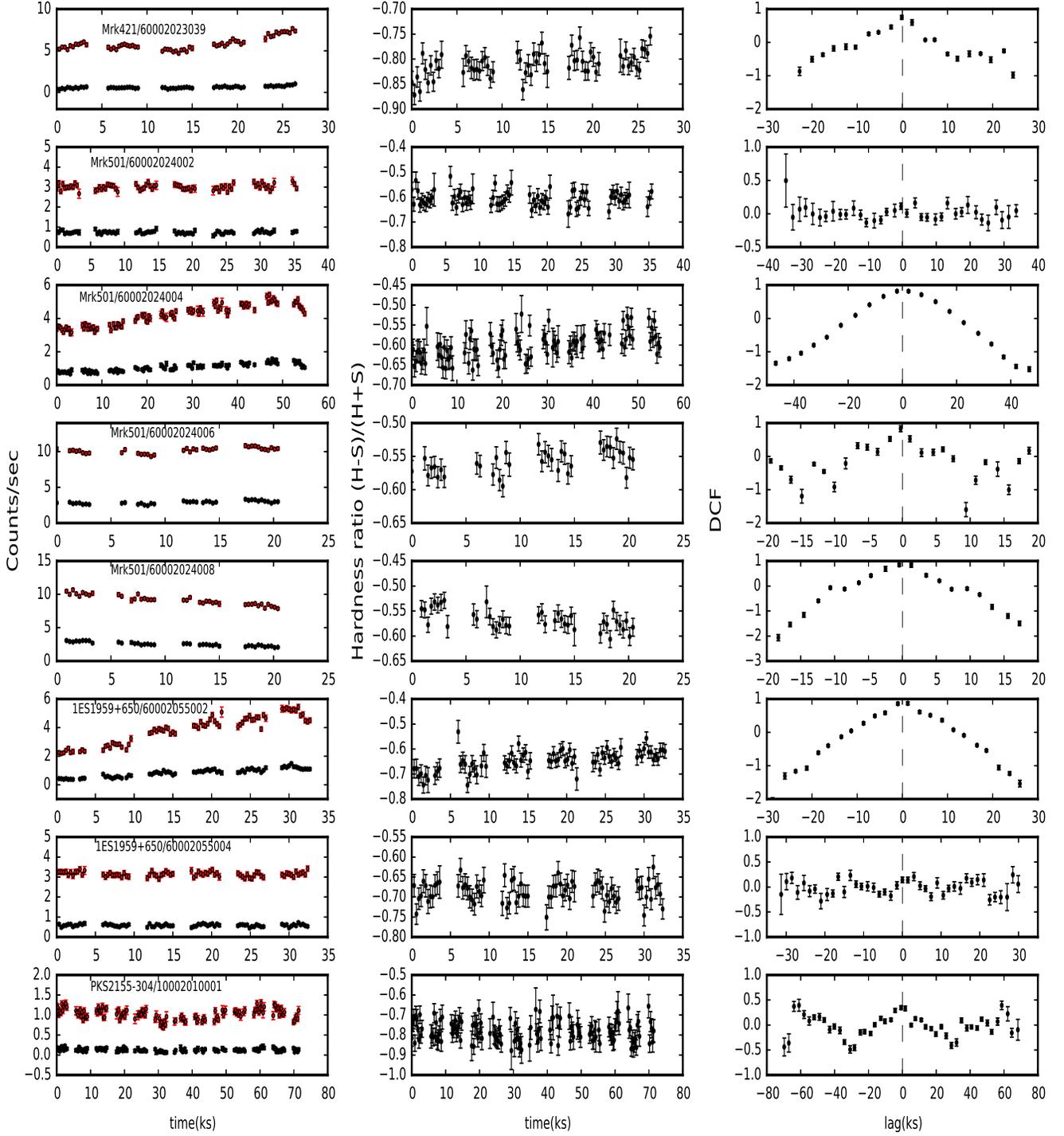}
\caption{\label{2_d} As in Figure \ref{2_a} for Mrk 421, Mrk 501, 1ES 1959$+$650 and PKS 2155$-$304.}
\end{figure*}

\begin{figure*}
\centering
\includegraphics[width=18cm, height=20cm]{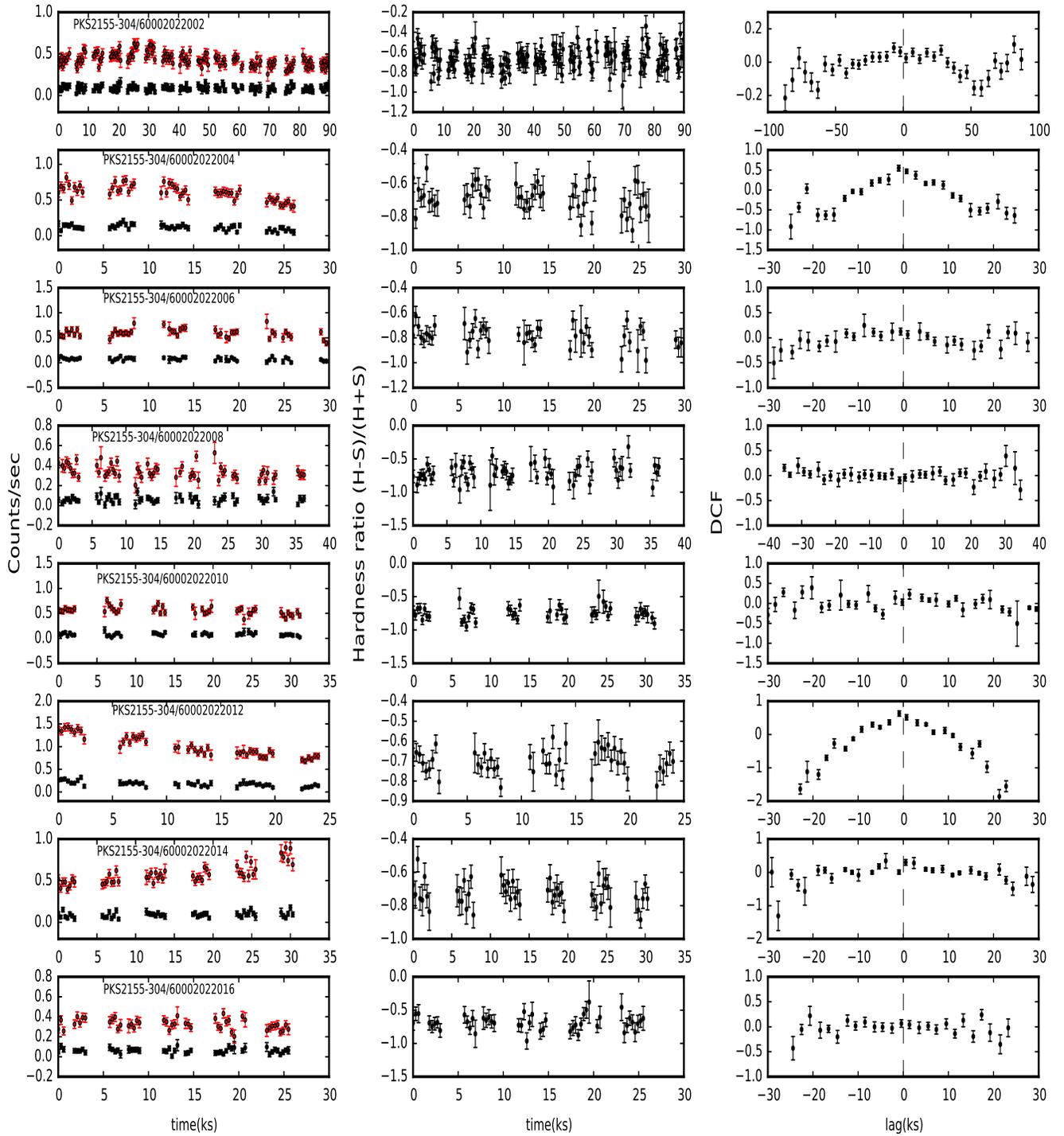}
\caption{\label{2_e} As in Figure \ref{2_a} for PKS 2155$-$304.}
\end{figure*}
\end{subfigures}

\begin{subfigures}
\label{fig:fig3}
\begin{figure*}
\centering
\includegraphics[width=18cm, height=20cm]{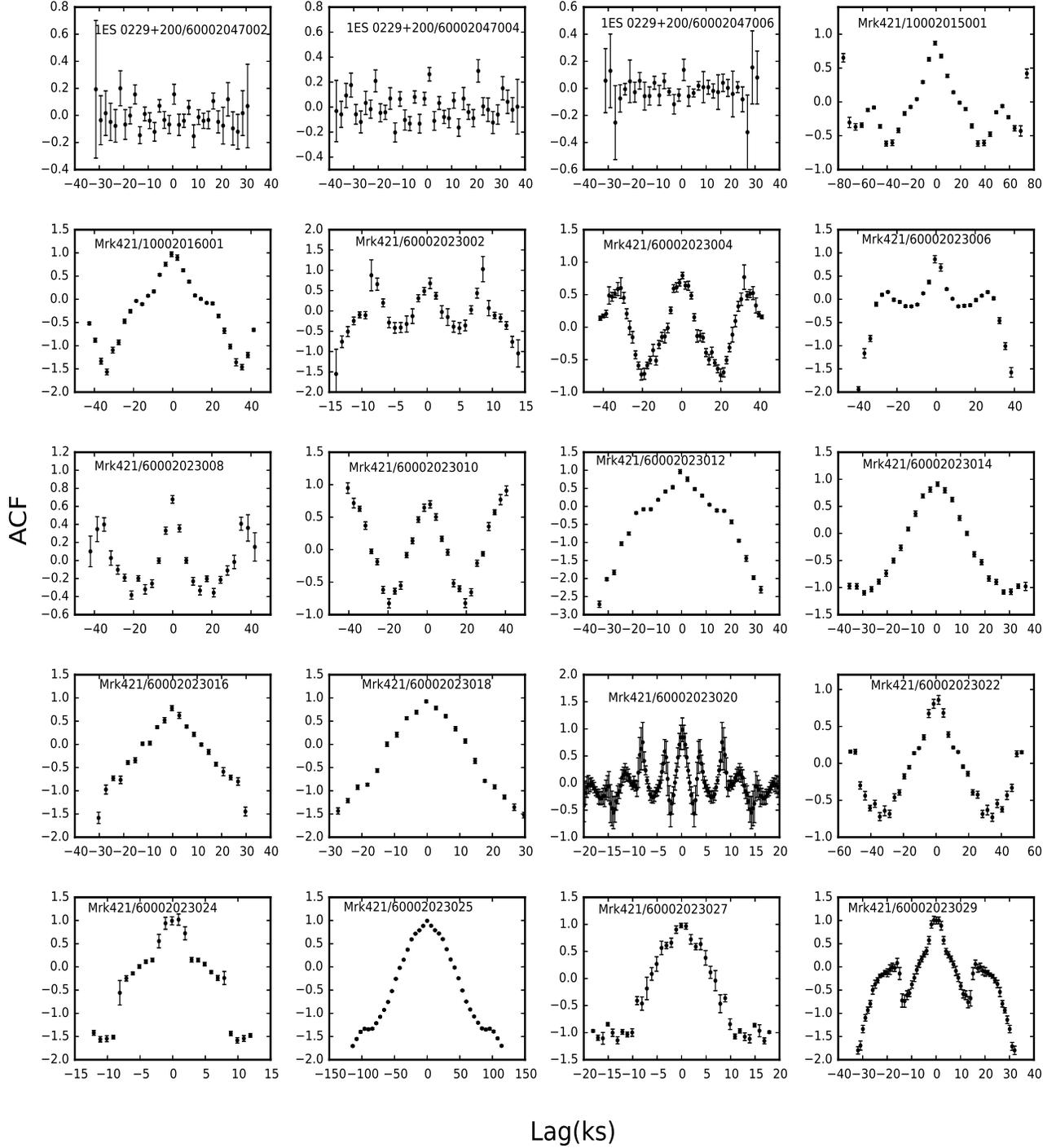}
\caption{\label{3_a} Auto Correlation functions for LCs of the blazars 1ES 0229$+$200 and Mrk 421.}
\end{figure*}

\begin{figure*}
\centering
\includegraphics[width=18cm, height=20cm]{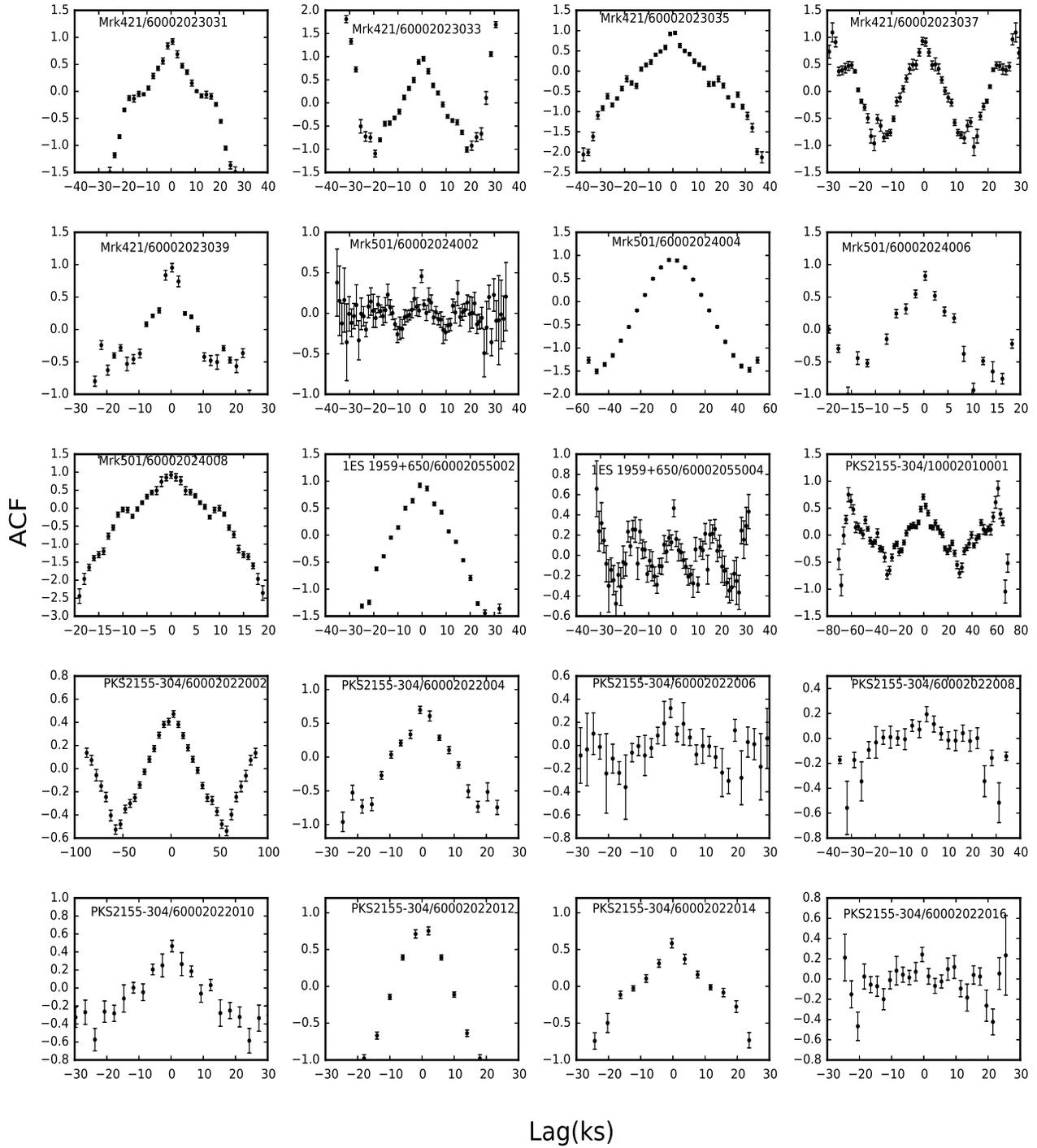}
\caption{\label{3_b} ACFs for the blazars Mrk 401, Mrk 501, 1ES 1959$+$650 and PKS 2155$-$304.}
\end{figure*}
\end{subfigures}


\section{RESULTS} \label{sec:result}

\begin{table*}
\centering
\caption{X-ray variability parameters.}
 \label{tab:var_par}
 \begin{tabular}{llcccccc}
  \hline
  Blazar Name  &  Obs. ID       &  \multicolumn{3}{c}{ $F_{var}(percent)$} & ACF(ks) & Bin-size(ks)\\
               &                 &  Soft (3-10 keV) & Hard (10-79 keV) & Total (3-79 keV) &		&	\\

  \hline
    
1ES 0229+200
& 60002047002 &         -             & $  15.632 \pm  4.757 $ &            -            &    -   & 2.00 \\
& 60002047004 &         -             &             -          &            -            &    -   & 2.00 \\
& 60002047006 & $ 8.638  \pm  2.557 $ &             -          &  $ 4.460  \pm   3.411 $ &    -   & 2.00 \\

MRK 421
& 10002015001 & $21.105  \pm  0.341 $ & $  26.899 \pm ~1.179 $ &  $21.217  \pm   0.260 $ &    -   & 5.00 \\
& 10002016001 & $31.225  \pm  0.337 $ & $  48.503 \pm ~1.227 $ &  $32.314  \pm   0.305 $ &    -   & 3.00 \\
& 60002023002 & $~7.531  \pm  1.183 $ & $  68.747 \pm ~8.125 $ &  $~9.159  \pm   1.029 $ &   4.9  & 0.90 \\
& 60002023004 & $12.173  \pm  0.874 $ & $  75.723 \pm 13.099 $ &  $11.485  \pm   0.860 $ &   19.9 & 1.50 \\
& 60002023006 & $20.051  \pm  0.402 $ & $  39.626 \pm  1.695 $ &  $21.595  \pm   0.392 $ &   11.4 & 3.00 \\
& 60002023008 & $13.566  \pm  0.721 $ & $  31.496 \pm  9.967 $ &  $10.275  \pm   0.757 $ &    -   & 3.50 \\
& 60002023010 & $10.371  \pm  0.517 $ & $  ~8.027 \pm  3.382 $ &  $10.264  \pm   0.411 $ &   19.5 & 3.00 \\
& 60002023012 & $20.310  \pm  0.463 $ & $  33.744 \pm  1.475 $ &  $21.556  \pm   0.424 $ &    -   & 3.00 \\
& 60002023014 & $25.203  \pm  1.219 $ & $  21.537 \pm 11.947 $ &  $26.702  \pm   0.652 $ &    -   & 3.00 \\
& 60002023016 & $12.497  \pm  0.550 $ & $  13.995 \pm  2.175 $ &  $11.815  \pm   0.421 $ &    -   & 3.00 \\
& 60002023018 & $14.707  \pm  0.453 $ & $  14.930 \pm  2.339 $ &  $14.583  \pm   0.420 $ &    -   & 3.00 \\
& 60002023020 & $10.217  \pm  1.284 $ &            -           &  $~9.164  \pm   0.374 $ &   2.5  & 0.30 \\
& 60002023022 & $22.052  \pm  0.386 $ & $  40.386 \pm  1.243 $ &  $24.265  \pm   0.193 $ &   32.8 & 3.00 \\
& 60002023024 & $15.181  \pm  0.431 $ & $  17.344 \pm  1.401 $ &  $15.458  \pm   0.411 $ &    -   & 1.00 \\
& 60002023025 & $59.837  \pm  0.131 $ & $  64.919 \pm  0.351 $ &  $60.497  \pm   0.123 $ &    -   & 6.00 \\
& 60002023027 & $13.406  \pm  0.196 $ & $  17.386 \pm  0.486 $ &  $13.945  \pm   0.182 $ &    -   & 1.00 \\
& 60002023029 & $25.285  \pm  0.184 $ & $  26.116 \pm  0.517 $ &  $22.266  \pm   0.175 $ &   13.1 & 1.00 \\
& 60002023031 & $31.141  \pm  0.108 $ & $  36.316 \pm  0.229 $ &  $32.073  \pm   0.098 $ &    -   & 2.00 \\
& 60002023033 & $19.344  \pm  0.540 $ & $  25.137 \pm  0.458 $ &  $18.747  \pm   0.169 $ &    -   & 2.00 \\
& 60002023035 & $39.043  \pm  0.178 $ & $  44.841 \pm  0.427 $ &  $39.943  \pm   0.165 $ &    -   & 2.00 \\
& 60002023037 & $17.584  \pm  0.404 $ & $  24.000 \pm  1.382 $ &  $18.268  \pm   0.387 $ &  12.6  & 1.00 \\
& 60002023039 & $12.310  \pm  0.460 $ & $  17.255 \pm  1.798 $ &  $12.773  \pm   0.443 $ &    -   & 2.00 \\

MRK 501
& 60002024002 & $ 1.327  \pm  1.406 $ &           -            &  $ 1.391  \pm   1.121 $ &    -   & 1.00 \\
& 60002024004 & $14.572  \pm  0.627 $ & $  20.901 \pm  1.940 $ &  $15.540  \pm   0.354 $ &    -   & 5.00 \\
& 60002024006 & $ 4.278  \pm  0.432 $ & $   5.138 \pm  0.925 $ &  $ 3.743  \pm   0.351 $ &    -   & 2.00 \\
& 60002024008 & $ 6.338  \pm  0.670 $ & $   9.067 \pm  1.400 $ &  $ 8.284  \pm   0.358 $ &   8.0  & 1.00 \\

1ES 1959+650
& 60002055002 & $26.849  \pm  1.192 $ & $  35.449 \pm  2.876 $ &  $26.629  \pm   0.451 $ &    -   & 3.00 \\
& 60002055004 &           -           & $   6.522 \pm  2.963 $ &            -            &    -   & 1.00 \\

PKS 2155-304
& 10002010001 & $12.901  \pm  0.915 $ & $  33.771 \pm  5.960 $ &  $ 10.601 \pm   0.744 $ &   29.6 & 2.00 \\
& 60002022002 &           -           & $  21.275 \pm  5.478 $ &  $ 11.020 \pm   1.109 $ &   57.4 & 5.00 \\
& 60002022004 & $12.551  \pm  2.437 $ & $  21.333 \pm  7.654 $ &  $ 14.931 \pm   1.400 $ &    -   & 3.00 \\
& 60002022006 & $10.830  \pm  4.268 $ &            -           &  $  9.256 \pm   2.162 $ &    -   & 2.00 \\
& 60002022008 &           -           & $  28.233 \pm 15.966 $ &  $ 13.504 \pm   3.000 $ &    -   & 3.00 \\
& 60002022010 &           -           & $  12.931 \pm 15.148 $ &            -            &    -   & 3.00 \\
& 60002022012 & $20.963  \pm  1.300 $ & $  24.613 \pm  3.958 $ &  $ 20.916 \pm   1.160 $ &    -   & 4.00 \\
& 60002022014 & $16.726  \pm  2.082 $ &            -           &  $ 16.992 \pm   1.554 $ &    -   & 4.00 \\
& 60002022016 & $16.810  \pm  2.876 $ &            -           &  $  3.298 \pm   8.251 $ &    -   & 2.00 \\
                                   
  \hline

   \end{tabular}
 \end{table*}

\subsection{1ES 0229$+$200} \label{subsec:0229}
The BL Lac object 1ES 0229$+$200 ($\alpha_{\rm 2000}$ = 02h32m53.2s; $\delta_{\rm 2000} = +20^{\circ}16^{\prime}21^{\prime\prime}$) 
is a HBL at $z=0.1396$ \citep{2005ApJ...631..762W}. 
Recently, \cite{2015arXiv150904470C} reported flux variability at VHE on monthly and yearly timescales by studying long-term observations 
of 1ES 0229$+$200 (from 2004 to 2013) with \textit{H.E.S.S.} and also found a hint of correlation between X-ray and 
VHE emissions. 

\par \textit{NuSTAR} observed 1ES 0229$+$200 for 16.26 ks, 20.29 ks and 18.02 ks on 2013 October 2, 5, and 10, respectively. 
These LCs and their ACFs are plotted in Figures 1a, and 3a, respectively. The count rates are low and the data noisy, so all 
fractional variances are consistent with no detectable variability on any of those days in the entire 3--79 keV energy band. 
But variations in hard and soft bands are nominally seen on 2013 October 2 and 10, respectively. Unsurprisingly, the ACFs 
show no hint of an IDV timescale in  the \textit{NuSTAR} range.

\par The soft and hard LCs (left panel), HR plots (middle panel) and the DCF plots between soft and hard band (right panel) 
of 1ES 0229$+$200 are shown in Fig.\ 2a. No significant spectral changes are seen from the HR plots in any of the three 
observations.  Naturally, given the lack of significant variations, the DCF plots are all flat and consistent with 0 throughout.

\subsection{Mrk 421} 
\label{subsec:421}
Markarian 421 (Mrk 421; $\alpha_{\rm 2000}$ = 11h04m19s; $\delta_{\rm 2000} = +38^{\circ}11^{\prime}41^{\prime\prime}$), is one of the nearest ($z=0.031$) BL Lac objects, and was the first extragalactic object detected at 
TeV energy \citep{1992Natur.358..477P}. It is classified as an HBL because its synchrotron peak lies at soft X-rays.
\par Mrk 421 is highly variable at all timescales over the entire electromagnetic spectrum and has been extensively studied 
during its flaring states (e.g. \cite{1994IAUC.5993....2T,1995ApJ...438L..59K,1995IAUC.6167....1T,1999ApJS..121..131F,2004A&A...422..505G,2004A&A...427..769T,2005A&A...440..409T,2008ATel.1574....1C,2008ATel.1583....1P,2008ICRC....3..973S,2012AJ....143...23G,2012A&A...545A.117L,2013A&A...559A..75B,2013EPJWC..6104020R,2014ApJ...782..110A}). 
In the optical region variations of 4.9 mag were found over the course of several years \citep{1976ARA&A..14..173S}. 
In the spring to summer of 2006, a major flare was recorded with the flux reaching up to $\sim 8.5$ mCrab in the 2.0-10.0 keV energy range 
(\cite{2009A&A...501..879T}; \cite{2009ApJ...699.1964U}). 
In 2010 January and February,  strong X-ray flares were detected, with the maximum flux recorded to be $120\pm10$ mCrab and $164\pm17$ mCrab 
respectively, with the latter being the largest ever reported from the source \citep{2015ApJ...798...27I}. In 2012 and 2013, Mrk 421 displayed 
two flares \citep{2015MNRAS.448.3121H} and the gamma-ray flare in 2012 was observed without a simultaneous X-ray flare, a behaviour 
called an `orphan flare' \citep{2015arXiv150801438F}. Due to its variable nature at all wavelengths and its proximity, it has been studied 
in several multi-wavelength observational campaigns (e.g. 
\cite{2005ApJ...630..130B,2008ApJ...677..906F,2008ChJAA...8..395G,2009ApJ...695..596H,2012AJ....143...23G,2012A&A...541A.140S,
2016arXiv160509017M,2016PASP..128g4101L}). Recently, \cite{2016A&A...591A..83S} reported on a long-term study of Mrk 421 with the High Altitude Gamma Ray (HAGAR) telescope array at Hanle, India, and found strong correlations between the gamma and radio wavelengths and between the optical and gamma wavebands, but saw no correlation between gamma and X-ray emissions. They also saw that the variability depends on energy, being maximum in X-ray and VHE bands.

\begin{figure}
\centering
\includegraphics[width=9cm, height=9cm]{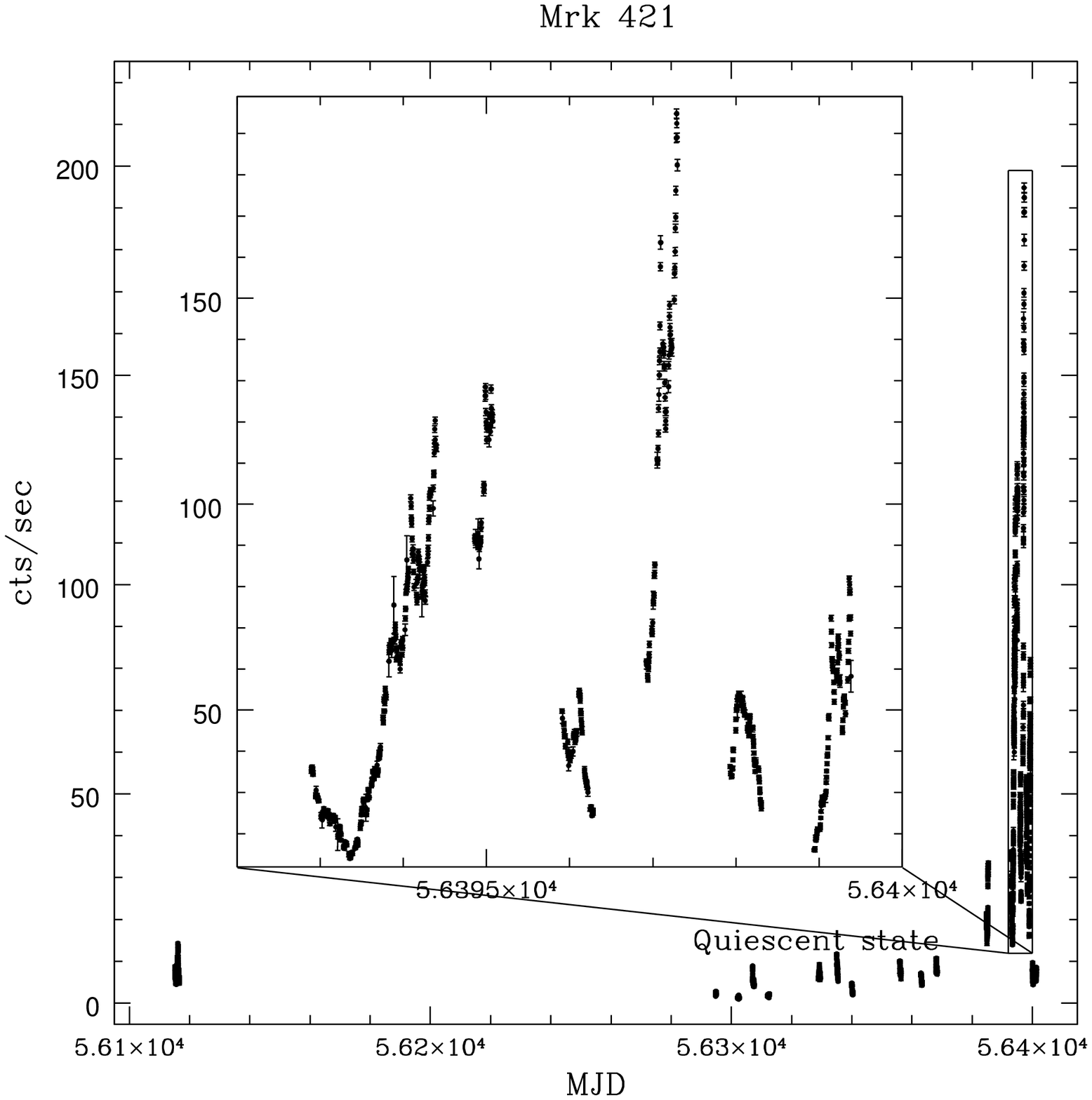}
\caption{\label{3}Short term variability of Mrk 421}
\end{figure}

\par Mrk 421 was first observed with \textit{NuSTAR} twice during  July 2012 as part of the calibration process for the telescope. It was then observed many times  in January -- April 2013 as part of an extensive and intensive 
multi-wavelength campaign that involved simultaneous or quasi-simultaneous data obtained in the radio, optical, soft X-ray, hard X-ray, and GeV through TeV gamma-ray bands \citep{2016ApJ...819..156B}.  The results of the first part of this study, covering the the low-flux state (in January -- March 2013) 
were reported by \cite{2016ApJ...819..156B} and \cite{2016arXiv160603659K}. 
 \cite{2015ApJ...811..143P} presented a variability analysis and \cite{2015A&A...580A.100S} presented analysis of the 
spectral variations during the strong flaring state  seen in April 2013 using this \textit{NuSTAR} data. 

The IDV LCs  of Mrk 421 that we have re-reduced and plotted in  Fig.\ 1 indicate that variations appear to  seen during every one of the 22 observations.  Our overall fractional variability amplitudes ($F_{var}$) given in Table 3 confirm these impressions   
and thus  echo the results of \cite{2016ApJ...819..156B} and \cite{2015ApJ...811..143P} in this regard. 
The fractional variability values for soft and hard bands, given in Table 3, indicate that the variability in the hard band is  stronger than that in the soft band for 19 out of 22 observations.
The complete LC obtained from our analysis of the \textit{NuSTAR} data from Mrk 421 is shown in Fig.\ 4. Very
strong flares were seen while the source was monitored nearly continuously during  2013 April 10 -- 16 \citep{2015ApJ...811..143P}. The 
 unprecedented outburst during this period is marked with a box in Fig.\ 4 and we have plotted a 
zoomed version of that box as an inset to the same figure. This is a double peaked outburst in which the first
flare appears to have a nearly a Gaussian shape with peak flux at $\sim$ MJD 56395 while the  second one, occurring
two days later, is even stronger, and evinces  a very sharp rise and decay.

\par The LCs in soft and hard bands (left panel), hardness ratio plots (middle panel) and the DCF plots (right panel) are shown in Figs.\ 2a, 2b, 2c, and 2d. The variations of hardness ratio with time show clear features of spectral changes that are stronger during flares. The spectra harden with the increasing flux, providing evidence for a general ``harder when brighter" behavior of blazars. Similar X-ray variability behavior was already seen in \textit{XMM-Newton} observations (e.g. \cite{2003A&A...402..929B,2004A&A...424..841R}) and in \textit{NuSTAR} observations (e.g. \cite{2015ApJ...811..143P,2016ApJ...819..156B}) of Mrk 421.  As can be seen from DCF plots, the soft and hard bands are positively correlated with zero time lag, indicating that their emissions  come from the same emitting region at the same time. 

\par The ACFs of Mrk 421 are plotted in Fig. \ref{fig:fig3}, and 8 of them show structures indicative of timescales. The dates on which observations began for which these structures were seen in the ACFs are 2013 January 2, 10 and 15, 2013 February 6, 2013 March 17, and 2013 April 2, 13 and 18, and the corresponding putative ``timescales", which range from 2.5 to 32.8 ks, are given in Table \ref{tab:var_par}. The other ACFs do not show any variability timescales or are too noisy to allow any detections of them. 
\subsection{Mrk 501} \label{subsec:501}

The TeV blazar Markarian 501 (Mrk 501;  $\alpha_{\rm 2000}$ = 16h53m52s; $\delta_{\rm 2000} =	+39^{\circ}45^{\prime}37^{\prime\prime}$), at $z= 0.034$, was the second extragalactic object detected at TeV energies. 
It was first detected at energy greater than 300 GeV by the Whipple Observatory \citep{1996ApJ...456L..83Q}.
In 1997 Mrk 501 went into a surprisingly high state with the flux recorded 
up to 10 Crab at energies \textgreater 1 TeV and it displayed strong VHE variability \citep{1997ApJ...487L.143C,1998ApJ...501L..17S,1997A&A...327L...5A,1999A&A...349...11A,1999A&A...342...69A}. 
On 16 April 1997 the highest VHE flux ($ F \sim 8.3 \times 10^{-10}$ cm$^{-2}$ s$^{-1}$ at energies \textgreater 250 GeV) ever was 
recorded \citep{1999A&A...350...17D}. Fast VHE flux variations with a flux doubling time of $\sim 2 $ minutes were observed from Mrk 501 
in July 2005 \citep{2007ApJ...669..862A}. \cite{2012A&A...541A..31N} reported an orphan VHE gamma-ray flare in 2009 which was not accompanied by an X-ray 
flare. In the optical band, Mrk 501 has also shown flux variability on different 
timescales (e.g. \cite{2008NewA...13..375G,2012NewA...17....8G,2016MNRAS.458.1127G,2016ApJS..222...24X}). Recently, in June 2014 
H.E.S.S. observed major flaring activity when the flux reached over 1 Crab and rapid flux variability was recorded at VHE 
($\sim$ 2--20 TeV) \citep{2015arXiv150904893C}. 

\par Mrk 501 was observed with \textit{NuSTAR} on four occasions between 2013 April 13 and 2013 July 13 as part of an extensive multi-wavelength campaign. Data was collected  from radio through optical, UV, X-ray and several gamma-ray bands, with the last two observations made as part of a target of opportunity program because of elevated states detected by other telescopes \citep{2015ApJ...812...65F}.
The mean fluxes varied significantly from observation to observation and the LCs in Fig.\ 1b show clear IDV during the last three nights of observations. This is confirmed by the substantial values of $F_{var}$ given in Table \ref{tab:var_par} for those observations. The $F_{var}$
values in the soft and hard bands suggest  greater variations in the hard band. 

\par The soft and hard LCs of Mrk 501 are plotted in the left panels of Fig.\ 2d. As seen from the hardness ratio plots in  the middle panels of that figure,  no significant spectral changes were seen during for first observation, while for the other three observations the spectra become harder with increasing flux and softer with decreasing flux. Such spectral behaviour was also reported in other X-ray observations (e.g.\ \cite{1998ApJ...492L..17P,2016A&A...594A..76A}).  The DCF plots shown in the right panels of Fig.\ 2d show a positive correlation with zero lag between the soft and hard bands, except for the first observation, during which there was negligible variability.  Hence for the first observation the ACF plot given in Fig.\ 3b is essentially noisy.  While  the second and third observations show clear ACFs, those data 
do not give any  indication of a  variability timescale. The ACF of the last observation, however,  does indicate a possible timescale of $\sim 8.0 $ ks.

\subsection{1ES 1959$+$650} \label{subsec:1959}

The HBL 1ES 1959+650 ($\alpha_{\rm 2000}$ = 19h59m59.8s; $\delta_{\rm 2000} = +65^{\circ}08^{\prime}55^{\prime\prime}$) is a BL Lac
object at $z=0.48$ \cite{1996ApJS..104..251P}). 
This blazar was first detected at X-rays with the \textit{Einstein} IPC Slew Survey \citep{1992ApJS...80..257E} and was
 further observed with \textit{ROSAT} in 1996 and with \textit{BeppoSAX} in 1997 \citep{2002A&A...383..410B}, as well as 
with \textit{RXTE-ARGOS}, \textit{Swift} and \textit{XMM-Newton} 
\citep{2002ApJ...571..763G,2003A&A...412..711T,2008A&A...478..395M}. 

\par 1ES 1959$+$650 was observed with \textit{NuSTAR} on 2014 September 17 and 22, with good exposure times of  19.61 ks and 20.34 ks, respectively. The LCs and ACFs   are shown in Figs.\ 1b and 3b, respectively. During the first observation the flux increased dramatically over the span of 30 ks.  The count rate was both lower and steadier during the second measurements. 
The ACF plot for the observation on 2014 September 17 does not show a timescale, while the observation on 2014 September 22 yields a noisy ACF and provides no useful information.
\par The soft and hard band LCs (left panels), hardness ratio plots (middle panels) and the DCF between the two energy bands (right panels) for this blazar are plotted in Fig.\ 2d. The hardness ratio plots reveal no spectral variations. The DCF plot for the observation on 2014 September 17 exhibits a correlation between the soft and hard bands with zero lag, while the lack of variability naturally means that no such correlation can be seen for the observation on 2014 September 22.

\subsection{PKS 2155$-$304} 
\label{subsec:2155}
The HBL PKS 2155$-$304  ($\alpha_{\rm 2000}$ = 21h58m52.7s; $\delta_{\rm 2000} =  -30^{\circ}13^{\prime}18^{\prime\prime};  
z = 0.116$, \cite{1993ApJ...411L..63F,2016MNRAS.455..618F}) is the brightest BL Lac object in UV to TeV energies in the southern 
hemisphere. 
It was  first recognized as a TeV blazar by the Durham MK6 telescopes \citep{1999ApJ...513..161C}. 
VHE flux variability on timescales of minutes was also found \citep{2007ApJ...664L..71A}. PKS 2155$-$304 has 
been observed at all wavelengths and flux variability on diverse timescales has been reported by many authors (e.g. 
\cite{1993ApJ...411..614U,2010ApJ...718..279G,2011JApA...32..155G,
2014RAA....14..933Z,2015arXiv150903104C,2016AJ....151...54S,
2016NewA...44...21B} and references therein). 
\cite{2009A&A...506L..17L} reported a possible $\sim 4.6$ hr quasi-periodic oscillation (QPO) in  an X-ray LC using \textit{XMM-Newton} data.

\par \textit{NuSTAR} observed PKS 2155$-$304 on nine occasions between 2012 July 8 and 2013 September 28 where the exposures ranged from 10.53 ks  
to 45.06 ks, as given in Table 1.  
All these LCs and ACFs  are plotted in Figures 1b and 3b, respectively. Although the count rates are low, a visual inspection of the LCs indicates that IDV appears to be present on at most dates. The $F_{var}$ values and their errors in 3--79 keV range given in Table \ref{tab:var_par} confirm that IDV was detected during 7 of the 9 observations. The $F_{var}$ values in soft and hard bands are also given in Table \ref{tab:var_par}.
The ACF plots for observations on 2012 July 8 and 2013 April 23 indicate variability timescales of 29.6 and 57.4 ks, respectively. The other ACF plots are noisy or do not show any timescale of variability. 
\par The soft and hard band LCs (left panel), hardness ratio plots (middle panel) and the DCF plots (right panel) of PKS 2155$-$304 are shown in Figs.\ 2d and  2e.   The hardness ratio plots are noisy, providing no useful information. The DCFs for observations on 2012 July 8, 2013 July 16, and 2013 August 26 show zero lag correlations between the bands, while for other observations no significant correlations are observed.

\section{DISCUSSION}
\label{sec:discussion}
\subsection{X-ray Flux Variability}
One of the important features of  blazar emissions at all measured wavelengths is rapid and strong flux variability on diverse timescales. By doing careful
blazar variability studies we can better understand the radiation mechanisms and also get information about the size, location and
structure of the emitting region (e.g.\ \cite{2003A&A...400..487C}). The theoretical models proposed to explain the totality of intrinsic AGN
variability can be coarsely classified as the relativistic-jet-based models (e.g. \cite{1985ApJ...298..114M,1992A&A...259..109G,2014ApJ...780...87M,2015JApA...36..255C} and the accretion disk based models
(e.g. \cite{1993ApJ...406..420M,1993ApJ...411..602C}). In blazars, the accretion disk radiation is almost always dominated by
the Doppler boosted radiation from the relativistic jets, so the accretion disk based models are generally not able to explain variability on any measurable timescales.
However, in radio-quiet quasars and  for blazars in very low states, it is possible that the IDV and STV can be explained by instabilities or hot spots on the accretion disk
(e.g. \cite{1993ApJ...406..420M,1993ApJ...411..602C}) because the accretion disk flux is not swamped by the
jet flux then.

\par The variations on LTV timescales (months to years) for blazars usually can be explained by shock-in-jet models (e.g.\ \cite{1985ApJ...298..114M,1995ARA&A..33..163W}), in which relativistic shocks, assumed to be formed by the disturbances in the inner
portion of the jet, propagate outward along the jet and produce major flux changes.  Motions of the shock through a helical jet (or helical structures within the jet) can cause
variations in  in the Doppler boosting by changing
the effective viewing angle and thus explain some LTV and STV (e.g., \cite{1992A&A...255...59C,1992A&A...259..109G}).
Smaller variations on STV and IDV timescales can be  explained
by turbulence behind the shock in the relativistic jet (e.g., \cite{2014ApJ...780...87M,2015JApA...36..255C,2016ApJ...820...12P}).

\par  In this work, we found  minimum hard X-ray variability timescales of 2.5, 8.0 and 29.6 ks for Mrk 421, Mrk 501 and PKS 2155$-$304, respectively. Such rapid X-ray variability timescales have been reported earlier in these sources \citep{2004ApJ...605..662C,2000ApJ...534L..39C,2000ApJ...528..243K}.  Extremely fast TeV variability timescales of a few minutes, as detected in PKS 2155$-$304 \citep{2007ApJ...664L..71A} and Mrk 501 \citep{2007ApJ...669..862A} are even shorter than the light crossing time of the Schwarzschild radius of the supermassive black holes at the centers of these blazars.
Such rapid timescales require the TeV flux emitting region to be very compact; however, the requirement that the TeV photons actually escape such a compact region without being absorbed via pair creation with the synchrotron photons implies the Lorentz factor ($\Gamma$) of the emitting region must be $\gtrsim 50$  (e.g.\cite{2006MNRAS.369.1287G,2008MNRAS.384L..19B}).There is very little other evidence for these extremely high bulk Lorentz factors in any AGN, and they appear to be  even more problematical for TeV blazars because  much lower values of $\Gamma$ in these blazars have been inferred from the rather slow apparent  motions of their radio knots made using very long baseline arrays \citep{2004ApJ...600..115P,2004ApJ...600..127G}. 

\par  Models that involve the radio and TeV emission emitting regions having substantially different properties can avoid this apparent contradiction.  For example, \cite{2008MNRAS.386L..28G} proposed the ``needle" model to explain fast TeV variability through the localized magneto-centrifugal acceleration of beams of electrons at small pitch angles along the magnetic field lines. Their model predicts TeV flares without simultaneous X-ray flares, and thus can explain the so called `orphan' flares.  However, during the 2013 April outburst  (Figs.\ 1 and 4) very fast ($\sim 14 $ minutes) hard X-ray variability is seen in the \textit{NuSTAR} LCs of Mrk 421 \citep{2015ApJ...811..143P}, accompanied by variations at VHE $\gamma-rays$ \citep{2013ATel.4976....1C}. So the ``needle" model fails to explain the spectacular 2013 April outburst of Mrk 421. Those flares can,  however, be explained with a ``jets-in-a-jet" model proposed by \cite{2009MNRAS.395L..29G}, which relies on  magnetic reconnection. This model predicts rapid TeV variations along with fast X-ray variability. In this model, the relativistic outflow of material from the magnetic reconnection regions gives rise to rapid flares through synchrotron-self-Compton emission. The model also has the flexibility to produce more slowly varying flares due to multiple reconnection regions or the tearing of a large reconnection site. The presence of more slowly varying short-term flares in the complete \textit{NuSTAR} LC of Mrk 421 (Fig.\ 4), lends some further support to this type of ``jets-in-a-jet" model, as discussed by Paliya et al.\ (2015).

\par Assuming that the hard X-ray emission from the HBLs is mainly due to synchrotron emission, certain  parameters can be estimated in a fashion that does not depend on the details of the acceleration models.  We initially follow Paliya et al.\ (2015)
and recall that the synchrotron cooling timescale, in the observer's frame, of an electron with energy $E = \gamma m_ec^2$ is \citep{2002ApJ...572..762Z}
\begin{equation}
t_{cool}(\gamma)  \simeq 7.74 \times 10^8 \frac{(1+z)}{\delta}B^{-2}\gamma^{-1}  {\rm s}, 
\end{equation}
where $\delta$ is the bulk Doppler factor, $B$ is the magnetic field strength in gauss, and $\gamma$ is the electron Lorentz factor.
For a given magnetic field strength and electron Lorentz factor the  frequency at which synchrotron emission occurs is, e.g., \citep{2015ApJ...811..143P},
\begin{equation}
\nu \equiv \nu_{19} \times10^{19}{\rm Hz} \simeq 4.2 \times 10^6 \frac{\delta}{1+z} B \gamma^2,
\end{equation}

\begin{table*}
\centering
\caption{Model parameters for \textit{NuSTAR} blazars}
 \label{tab:para}
 \begin{tabular}{lcccccc}
  \hline
Blazar  & $t_{var}$(s) & $\delta$ & $B$ (G) & $ \gamma    $  & $R$ (cm)  \\
\hline
Mrk 421 & ~2500  & 25 & $ \geq 0.12 $  & $\leq 9.0 \times 10^5 $ & $ \leq 1.8 \times 10^{15} $\\
Mrk 501 & ~8000  & 15 & $ \geq 0.07 $  & $ \leq 1.5 \times 10^6 $ & $ \leq 3.5 \times 10^{15} $\\
PKS 2155$-$304  & 29600 & 30 & $ \geq 0.02 $  & $ \leq 2.0 \times 10^6 $ & $ \leq 2.4 \times 10^{16} $ \\
\hline
\end{tabular}
\end{table*}

where $0.08< \nu_{19} < 2$  for X-rays in the \textit{NuSTAR} band.  Combining these two equations with the physical requirement that the observed minimum variability timescale has to be larger than or equal to the synchrotron cooling
timescale, we get for Mrk 421 (with our minimum $t_{var} = 2500$ s and $z = 0.031$),
\begin{equation}
B \geq 0.35~  \delta^{-1/3} \nu_{19}^{-1/3}  {\rm G}.
\end{equation}
We note that this expression has different dependences on $\delta$ and $\nu_{19}$ than does the one (Eq.\ 5) in Paliya et al.\ (2015). 
For $\delta = 25$, we find $B \geq 0.12 \nu_{19}^{-1/3} $ G, which is close to the typical value of $ B \sim 0.1 $ G that is inferred from the SED modeling of Mrk 421. Assuming $\delta = 25$ and $B \geq 0.12 $ G, we can constrain the electron Lorentz factor to
\begin{equation}
\gamma \leq 9 \times 10^5 \nu_{19}^{1/2} .
\end{equation}
We can also estimate the characteristic size of the emitting region as
\begin{equation}
R \leq c t_{var}  \delta /(1+z) \leq 1.8 \times 10^{15} cm. 
\end{equation}

\par Similarly, we can constrain these parameters for Mrk 501 and PKS 2155$-$304, where we also have reasonable observed timescales from \textit{NuSTAR} data, where typical values for $\delta$ are used  \citep{2015ApJ...812...65F,2000ApJ...528..243K} and where $\nu_{19} =1$ is assumed. These are given inTable \ref{tab:para}.
\par The rapid hard X-ray variability observed by \textit{NuSTAR} suggests that the relativistic electrons responsible for the hard X-ray emission of TeV HBLs must be repeatedly injected (accelerated), since the high energy electrons have short cooling timescales. The relativistic electrons are known to be  accelerated at shock fronts within jets, although additional acceleration regions are possible.  Diffusive shock acceleration (e.g.\ \cite{1987PhR...154....1B})  is an  electron acceleration mechanism which could be responsible for both the flux variations and the observed spectral hardening at high energies.
 
\subsection{X-ray Spectral Variability}
We examined the X-ray spectral variability of five TeV HBLs in this work using a model independent hardness ratio analysis. Although this method does not provide any direct information about the physical parameters responsible for spectral changes it is  the simplest way to study the spectral  variability. Given the steepness of the X-ray spectrum, which means that that the number of hard counts is significantly lower than that of the soft counts (see Fig.\ 2) it is difficult to do a more detailed spectral analysis. For two bright HSP blazars, Mrk 421 and Mrk 501, we found that the hardness ratio increases with increasing count rates, that is, their spectra become flatter with increasing flux. Such a ``hardening when brightening" trend appears to be a general feature of HSP type blazars, as it  was also noticed in earlier X-ray observations (e.g. \cite{1998ApJ...492L..17P,2002ApJ...572..762Z,2003A&A...402..929B,2004A&A...424..841R}). In the case of Mrk 421, we observed that the variations in hardness ratio values with time are particularly large during flares.

\section{CONCLUSIONS}
\label{conclusion}
We examined   the 40 longest X-ray LCs of  the five TeV HBLs  that \textit{NuSTAR} has observed for intraday variability and also searched for possible timescales of variability using  discrete autocorrelation analyses.   The variability in the hard X-ray emission that we have investigated here presumably originates in compact regions within the jet.  The vast majority of the observations of these five TeV blazars that have decent count rates show significant IDV in the \textit{NuSTAR} band, and we have found strong evidence for IDV for:  Mrk 421, in all 22 of 22 LCs; Mrk 501, in 3 of 4; 1ES 1959$+$650, in 1 of 2; and in PKS 2155$-$304, in 7 of 9.

\par Using ACFs, we found evidence for timescales ranging from 2.5 ks to 32.8 ks in eight LCs of Mrk 421, a timescale of 8.0 ks for one LC of Mrk 501, and 29.6 and 57.4 ks timescales for two LCs of PKS 2155$-$304.
For another 29 LCs (3 of 1ES 0229+200, 14 of Mrk 421, 3 of Mrk 501, 2 of 1ES 1959$+$650 and 7 of PKS 2155$-$304), either the ACF plot is noisy or the data is good but no timescale of variability is clearly  present. Using the shortest observed variability timescales, we  estimated the values of magnetic field ($B$), electron Lorentz factor ($\gamma$) and size ($R$) of these emitting regions for Mrk 421, Mrk 501 and PKS 2155$-$304 (Table \ref{tab:para}).

\par We also employed a hardness ratio analysis to make a preliminary study of the X-ray spectral variability of these five TeV HBLs.  We found that the X-ray spectra harden with increasing count rates for Mrk 421 and Mrk 501.
Using a DCF analysis we performed a correlation study between soft (3--10 keV) and hard (10--79 keV) bands.  We found overall positive correlations with zero lag for Mrk 421 (in all 22 observations), for Mrk 501 (in 3 of 4), for 1ES 1959$+$650 (in 1 of 2) and for PKS 2155$-$304 (in 3 of 9).  These measurements indicate that the hard and soft X-ray emissions from these blazers are produced by the same populations of electrons.

We thank the anonymous referee for important comments, which helped us to improve the manuscript.
ACG is partially supported by the Chinese Academy of Sciences (CAS) President's International Fellowship Initiative (PIFI) (grant no.\ 2016VMB073).  PJW is grateful for hospitality at KIPAC, Stanford University, during a sabbatical.
This research has made use of data obtained with \textit{NuSTAR}, the first focusing hard X-ray mission managed by the Jet Propulsion Laboratory (JPL), and funded by the National Aeronautics and Space Administration (NASA). This research has also made use of the NuSTAR Data Analysis Software (NuSTARDAS) jointly developed by the ASI Science Data Center (ASDC, Italy) and the California Institute of Technology (Caltech, USA). 
\bibliographystyle{aasjournal}
\bibliography{master}

\begin{thebibliography}{}
\expandafter\ifx\csname natexlab\endcsname\relax\def\natexlab#1{#1}\fi

\bibitem[{{Abdo} {et~al.}(2010){Abdo}, {Ackermann}, {Agudo}, {Ajello}, {Aller},
  {Aller}, {Angelakis}, {Arkharov}, {Axelsson}, {Bach}, \&
  et~al.}]{2010ApJ...716...30A}
{Abdo}, A.~A., {Ackermann}, M., {Agudo}, I., {et~al.} 2010, \apj, 716, 30

\bibitem[{{Abdo} {et~al.}(2014){Abdo}, {Abeysekara}, {Allen}, {Aune}, {Barber},
  {Berley}, {Braun}, {Chen}, {Christopher}, {Delay}, {DeYoung}, {Dingus},
  {Ellsworth}, {Fraija}, {Gonz{\'a}lez}, {Goodman}, {Hays}, {Hoffman},
  {H{\"u}ntemeyer}, {Imran}, {Kolterman}, {Linnemann}, {Marinelli}, {McEnery},
  {Morgan}, {Mincer}, {Nemethy}, {Patricelli}, {Pretz}, {Ryan}, {Saz
  Parkinson}, {Schneider}, {Shoup}, {Sinnis}, {Smith}, {Vasileiou}, {Walker},
  {Williams}, \& {Yodh}}]{2014ApJ...782..110A}
{Abdo}, A.~A., {Abeysekara}, A.~U., {Allen}, B.~T., {et~al.} 2014, \apj, 782,
  110

\bibitem[{{Aharonian} {et~al.}(2007){Aharonian}, {Akhperjanian}, {Bazer-Bachi},
  {Behera}, {Beilicke}, \& {Benbow}}]{2007ApJ...664L..71A}
{Aharonian}, F., {Akhperjanian}, A.~G., {Bazer-Bachi}, A.~R., {et~al.} 2007,
  \apjl, 664, L71

\bibitem[{{Aharonian} {et~al.}(1997){Aharonian}, {Akhperjanian}, {Barrio},
  {Bernloehr}, {Beteta}, {Bradbury}, {Contreras}, {Cortina}, {Daum}, {Deckers},
  {Feigl}, {Fernandez}, {Fonseca}, {Frass}, {Funk}, {Gonzalez}, {Haustein},
  {Heinzelmann}, {Hemberger}, {Hermann}, {Hess}, {Heusler}, {Hofmann}, {Holl},
  {Horns}, {Kankanian}, {Kirstein}, {Koehler}, {Konopelko}, {Kornmayer},
  {Kranich}, {Krawczynski}, {Lampeitl}, {Lindner}, {Lorenz}, {Magnussen},
  {Meyer}, {Mirzoyan}, {Moeller}, {Moralejo}, {Padilla}, {Panter}, {Petry},
  {Plaga}, {Prahl}, {Prosch}, {Puehlhofer}, {Rauterberg}, {Rhode}, {Rivero},
  {Roehring}, {Sahakian}, {Samorski}, {Sanchez}, {Schmele}, {Schmidt}, {Stamm},
  {Ulrich}, {Voelk}, {Westerhoff}, {Wiebel-Sooth}, {Wiedner}, {Willmer}, \&
  {Wirth}}]{1997A&A...327L...5A}
{Aharonian}, F., {Akhperjanian}, A.~G., {Barrio}, J.~A., {et~al.} 1997, \aap,
  327, L5

\bibitem[{{Aharonian} {et~al.}(1999{\natexlab{a}}){Aharonian}, {Akhperjanian},
  {Barrio}, {Bernl{\"o}hr}, \& {Bojahr}}]{1999A&A...349...11A}
{Aharonian}, F.~A., {Akhperjanian}, A.~G., {Barrio}, J.~A., {Bernl{\"o}hr}, K.,
  \& {Bojahr}, H. 1999{\natexlab{a}}, \aap, 349, 11

\bibitem[{{Aharonian} {et~al.}(1999{\natexlab{b}}){Aharonian}, {Akhperjanian},
  {Barrio}, {Bernl{\"o}hr}, {Bojahr}, {Contreras}, {Cortina}, {Daum},
  {Deckers}, {Fonseca}, {Gonzalez}, {Heinzelmann}, {Hemberger}, {Hermann},
  {He{\ss}}, {Heusler}, {Hofmann}, {Hohl}, {Horns}, {Ibarra}, {Kankanyan},
  {Kirstein}, {K{\"o}hler}, {Konopelko}, {Kornmeyer}, {Kranich}, {Krawczynski},
  {Lampeitl}, {Lindner}, {Lorenz}, {Magnussen}, {Meyer}, {Mirzoyan},
  {Moralejo}, {Padilla}, {Panter}, {Petry}, {Plaga}, {Plyasheshnikov}, {Prahl},
  {P{\"u}hlhofer}, {Rauterberg}, {Renault}, {Rhode}, {Sahakian}, {Samorski},
  {Schmele}, {Schr{\"o}der}, {Stamm}, {V{\"o}lk}, {Wiebel-Sooth}, {Wiedner},
  {Willmer}, \& {Wirth}}]{1999A&A...342...69A}
{Aharonian}, F.~A., {Akhperjanian}, A.~G., {Barrio}, J.~A., {et~al.}
  1999{\natexlab{b}}, \aap, 342, 69

\bibitem[{{Albert} {et~al.}(2007){Albert}, {Aliu}, {Anderhub}, {Antoranz},
  {Armada}, {Baixeras}, {Barrio}, {Bartko}, {Bastieri}, {Becker}, {Bednarek},
  {Berger}, {Bigongiari}, {Biland}, {Bock}, {Bordas}, {Bosch-Ramon}, {Bretz},
  {Britvitch}, {Camara}, {Carmona}, {Chilingarian}, {Coarasa}, {Commichau},
  {Contreras}, {Cortina}, {Costado}, {Curtef}, {Danielyan}, {Dazzi}, {De
  Angelis}, {Delgado}, {de los Reyes}, {De Lotto}, {Domingo-Santamar{\'{\i}}a},
  {Dorner}, {Doro}, {Errando}, {Fagiolini}, {Ferenc}, {Fern{\'a}ndez}, {Firpo},
  {Flix}, {Fonseca}, {Font}, {Fuchs}, {Galante}, {Garc{\'{\i}}a-L{\'o}pez},
  {Garczarczyk}, {Gaug}, {Giller}, {Goebel}, {Hakobyan}, {Hayashida},
  {Hengstebeck}, {Herrero}, {H{\"o}hne}, {Hose}, {Hrupec}, {Hsu}, {Jacon},
  {Jogler}, {Kosyra}, {Kranich}, {Kritzer}, {Laille}, {Lindfors}, {Lombardi},
  {Longo}, {L{\'o}pez}, {L{\'o}pez}, {Lorenz}, {Majumdar}, {Maneva},
  {Mannheim}, {Mansutti}, {Mariotti}, {Mart{\'{\i}}nez}, {Mazin}, {Merck},
  {Meucci}, {Meyer}, {Miranda}, {Mirzoyan}, {Mizobuchi}, {Moralejo}, {Nieto},
  {Nilsson}, {Ninkovic}, {O{\~n}a-Wilhelmi}, {Otte}, {Oya}, {Paneque},
  {Panniello}, {Paoletti}, {Paredes}, {Pasanen}, {Pascoli}, {Pauss}, {Pegna},
  {Persic}, {Peruzzo}, {Piccioli}, {Prandini}, {Puchades}, {Raymers}, {Rhode},
  {Rib{\'o}}, {Rico}, {Rissi}, {Robert}, {R{\"u}gamer}, {Saggion}, {Saito},
  {S{\'a}nchez}, {Sartori}, {Scalzotto}, {Scapin}, {Schmitt}, {Schweizer},
  {Shayduk}, {Shinozaki}, {Shore}, {Sidro}, {Sillanp{\"a}{\"a}}, {Sobczynska},
  {Stamerra}, {Stark}, {Takalo}, {Tavecchio}, {Temnikov}, {Tescaro}, {Teshima},
  {Torres}, {Turini}, {Vankov}, {Vitale}, {Wagner}, {Wibig}, {Wittek},
  {Zandanel}, {Zanin}, \& {Zapatero}}]{2007ApJ...669..862A}
{Albert}, J., {Aliu}, E., {Anderhub}, H., {et~al.} 2007, \apj, 669, 862

\bibitem[{{Aleksi{\'c}} {et~al.}(2015){Aleksi{\'c}}, {Ansoldi}, {Antonelli},
  {Antoranz}, {Babic}, {Bangale}, {Barres de Almeida}, {Barrio}, {Becerra
  Gonz{\'a}lez}, {Bednarek}, \& et~al.}]{2012A&A...541A.140S}
{Aleksi{\'c}}, J., {Ansoldi}, S., {Antonelli}, L.~A., {et~al.} 2015, \aap, 576,
  A126

\bibitem[{{Aliu} {et~al.}(2016){Aliu}, {Archambault}, {Archer}, {Arlen},
  {Aune}, {Barnacka}, {Behera}, {Beilicke}, {Benbow}, {Berger}, {Bird},
  {B{\"o}ttcher}, {Bouvier}, {Buchovecky}, {Buckley}, {Bugaev}, {Cardenzana},
  {Cerruti}, {Cesarini}, {Chen}, {Ciupik}, {Collins-Hughes}, {Connolly}, {Cui},
  {Dumm}, {Eisch}, {Falcone}, {Federici}, {Feng}, {Finley}, {Fleischhack},
  {Fortin}, {Fortson}, {Furniss}, {Galante}, {Gall}, {Gillanders}, {Griffin},
  {Griffiths}, {Grube}, {Gyuk}, {H{\"u}tten}, {H{\aa}kansson}, {Holder},
  {Hughes}, {Humensky}, {Johnson}, {Kaaret}, {Kar}, {Kelley-Hoskins},
  {Kertzman}, {Khassen}, {Kieda}, {Krause}, {Krawczynski}, {Krennrich}, {Lang},
  {Madhavan}, {Maier}, {McArthur}, {McCann}, {Meagher}, {Millis}, {Moriarty},
  {Mukherjee}, {Nieto}, {O'Faol{\'a}in de Bhr{\'o}ithe}, {Ong}, {Orr}, {Otte},
  {Pandel}, {Park}, {Pelassa}, {Perkins}, {Pichel}, {Pohl}, {Popkow}, {Quinn},
  {Ragan}, {Reyes}, {Reynolds}, {Roache}, {Rousselle}, {Rovero}, {Saxon},
  {Sembroski}, {Shahinyan}, {Sheidaei}, {Skole}, {Smith}, {Staszak},
  {Telezhinsky}, {Theiling}, {Todd}, {Tucci}, {Tyler}, {Varlotta}, {Vassiliev},
  {Vincent}, {Wakely}, {Weiner}, {Weinstein}, {Welsing}, {Wilhelm}, {Williams},
  \& {Zitzer}}]{2016A&A...594A..76A}
{Aliu}, E., {Archambault}, S., {Archer}, A., {et~al.} 2016, \aap, 594, A76

\bibitem[{{Balokovi{\'c}} {et~al.}(2016){Balokovi{\'c}}, {Paneque}, {Madejski},
  {Furniss}, {Chiang}, {Ajello}, {Alexander}, {Barret}, {Blandford}, {Boggs},
  \& et~al.}]{2016ApJ...819..156B}
{Balokovi{\'c}}, M., {Paneque}, D., {Madejski}, G., {et~al.} 2016, \apj, 819,
  156

\bibitem[{{Beckmann} {et~al.}(2002){Beckmann}, {Wolter}, {Celotti},
  {Costamante}, {Ghisellini}, {Maccacaro}, \&
  {Tagliaferri}}]{2002A&A...383..410B}
{Beckmann}, V., {Wolter}, A., {Celotti}, A., {et~al.} 2002, \aap, 383, 410

\bibitem[{{Begelman} {et~al.}(2008){Begelman}, {Fabian}, \&
  {Rees}}]{2008MNRAS.384L..19B}
{Begelman}, M.~C., {Fabian}, A.~C., \& {Rees}, M.~J. 2008, \mnras, 384, L19

\bibitem[{{Bhagwan} {et~al.}(2016){Bhagwan}, {Gupta}, {Papadakis}, \&
  {Wiita}}]{2016NewA...44...21B}
{Bhagwan}, J., {Gupta}, A.~C., {Papadakis}, I.~E., \& {Wiita}, P.~J. 2016, \na,
  44, 21

\bibitem[{{Blandford} \& {Eichler}(1987)}]{1987PhR...154....1B}
{Blandford}, R., \& {Eichler}, D. 1987, \physrep, 154, 1

\bibitem[{{Blasi} {et~al.}(2013){Blasi}, {Lico}, {Giroletti}, {Orienti},
  {Giovannini}, {Cotton}, {Edwards}, {Fuhrmann}, {Krichbaum}, {Kovalev},
  {Jorstad}, {Marscher}, {Kino}, {Paneque}, {Perez-Torres}, {Piner}, \&
  {Sokolovsky}}]{2013A&A...559A..75B}
{Blasi}, M.~G., {Lico}, R., {Giroletti}, M., {et~al.} 2013, \aap, 559, A75

\bibitem[{{B{\l}a{\.z}ejowski} {et~al.}(2005){B{\l}a{\.z}ejowski}, {Blaylock},
  {Bond}, {Bradbury}, {Buckley}, {Carter-Lewis}, {Celik}, {Cogan}, {Cui},
  {Daniel}, {Duke}, {Falcone}, {Fegan}, {Fegan}, {Finley}, {Fortson},
  {Gammell}, {Gibbs}, {Gillanders}, {Grube}, {Gutierrez}, {Hall}, {Hanna},
  {Holder}, {Horan}, {Humensky}, {Kenny}, {Kertzman}, {Kieda}, {Kildea},
  {Knapp}, {Kosack}, {Krawczynski}, {Krennrich}, {Lang}, {LeBohec}, {Linton},
  {Lloyd-Evans}, {Maier}, {Mendoza}, {Milovanovic}, {Moriarty}, {Nagai}, {Ong},
  {Power-Mooney}, {Quinn}, {Quinn}, {Ragan}, {Reynolds}, {Rebillot}, {Rose},
  {Schroedter}, {Sembroski}, {Swordy}, {Syson}, {Valcarel}, {Vassiliev},
  {Wakely}, {Walker}, {Weekes}, {White}, {Zweerink}, {VERITAS Collaboration},
  {Mochejska}, {Smith}, {Aller}, {Aller}, {Ter{\"a}sranta}, {Boltwood},
  {Sadun}, {Stanek}, {Adams}, {Foster}, {Hartman}, {Lai}, {B{\"o}ttcher},
  {Reimer}, \& {Jung}}]{2005ApJ...630..130B}
{B{\l}a{\.z}ejowski}, M., {Blaylock}, G., {Bond}, I.~H., {et~al.} 2005, \apj,
  630, 130

\bibitem[{{B{\"o}ttcher}(2007)}]{2007Ap&SS.307...69B}
{B{\"o}ttcher}, M. 2007, \apss, 307, 69

\bibitem[{{Brinkmann} {et~al.}(2003){Brinkmann}, {Papadakis}, {den Herder}, \&
  {Haberl}}]{2003A&A...402..929B}
{Brinkmann}, W., {Papadakis}, I.~E., {den Herder}, J.~W.~A., \& {Haberl}, F.
  2003, \aap, 402, 929

\bibitem[{{Calafut} \& {Wiita}(2015)}]{2015JApA...36..255C}
{Calafut}, V., \& {Wiita}, P.~J. 2015, Journal of Astrophysics and Astronomy,
  36, 255

\bibitem[{{Camenzind} \& {Krockenberger}(1992)}]{1992A&A...255...59C}
{Camenzind}, M., \& {Krockenberger}, M. 1992, \aap, 255, 59

\bibitem[{{Catanese} \& {Sambruna}(2000)}]{2000ApJ...534L..39C}
{Catanese}, M., \& {Sambruna}, R.~M. 2000, \apjl, 534, L39

\bibitem[{{Catanese} {et~al.}(1997){Catanese}, {Bradbury}, {Breslin},
  {Buckley}, {Carter-Lewis}, {Cawley}, {Dermer}, {Fegan}, {Finley}, {Gaidos},
  {Hillas}, {Johnson}, {Krennrich}, {Lamb}, {Lessard}, {Macomb}, {McEnery},
  {Moriarty}, {Quinn}, {Rodgers}, {Rose}, {Samuelson}, {Sembroski},
  {Srinivasan}, {Weekes}, \& {Zweerink}}]{1997ApJ...487L.143C}
{Catanese}, M., {Bradbury}, S.~M., {Breslin}, A.~C., {et~al.} 1997, \apjl, 487,
  L143

\bibitem[{{Chadwick} {et~al.}(1999){Chadwick}, {Lyons}, {McComb}, {Orford},
  {Osborne}, {Rayner}, {Shaw}, {Turver}, \& {Wieczorek}}]{1999ApJ...513..161C}
{Chadwick}, P.~M., {Lyons}, K., {McComb}, T.~J.~L., {et~al.} 1999, \apj, 513,
  161

\bibitem[{{Chakrabarti} \& {Wiita}(1993)}]{1993ApJ...411..602C}
{Chakrabarti}, S.~K., \& {Wiita}, P.~J. 1993, \apj, 411, 602

\bibitem[{{Chakraborty} {et~al.}(2015){Chakraborty}, {Cologna}, {Kastendieck},
  {Rieger}, {Romoli}, {Wagner}, {Jacholkowska}, {Taylor}, \& {for the
  H.~E.~S.~S.~Collaboration}}]{2015arXiv150904893C}
{Chakraborty}, N., {Cologna}, G., {Kastendieck}, M.~A., {et~al.} 2015, ArXiv
  e-prints, arXiv:1509.04893

\bibitem[{{Chevalier} {et~al.}(2015){Chevalier}, {Kastendieck}, {Rieger},
  {Maurin}, {Lenain}, \& {Giovanni Lamanna for the
  H.~E.~S.~S.~collaboration}}]{2015arXiv150903104C}
{Chevalier}, J., {Kastendieck}, M.~A., {Rieger}, F., {et~al.} 2015, ArXiv
  e-prints, arXiv:1509.03104

\bibitem[{{Ciprini} {et~al.}(2003){Ciprini}, {Tosti}, {Raiteri}, {Villata},
  {Ibrahimov}, {Nucciarelli}, \& {Lanteri}}]{2003A&A...400..487C}
{Ciprini}, S., {Tosti}, G., {Raiteri}, C.~M., {et~al.} 2003, \aap, 400, 487

\bibitem[{{Cologna} {et~al.}(2015){Cologna}, {Mohamed}, {Wagner},
  {Wierzcholska}, {Romoli}, {for the H.~E.~S.~S.~Collaboration}, \&
  {Kurtanidze}}]{2015arXiv150904470C}
{Cologna}, G., {Mohamed}, M., {Wagner}, S.~J., {et~al.} 2015, ArXiv e-prints,
  arXiv:1509.04470

\bibitem[{{Cortina} \& {Holder}(2013)}]{2013ATel.4976....1C}
{Cortina}, J., \& {Holder}, J. 2013, The Astronomer's Telegram, 4976

\bibitem[{{Costa} {et~al.}(2008){Costa}, {Del Monte}, {Donnarumma},
  {Evangelista}, {Feroci}, {Lapshov}, {Lazzarotto}, {Pacciani}, {Rapisarda},
  {Soffitta}, {Argan}, {Trois}, {Tavani}, {Pucella}, {D'Ammando}, {Vittorini},
  {Chen}, {Vercellone}, {Giuliani}, {Mereghetti}, {Pellizzoni}, {Perotti},
  {Fornari}, {Fiorini}, {Caraveo}, {Bulgarelli}, {Gianotti}, {Trifoglio}, {Di
  Cocco}, {Labanti}, {Fuschino}, {Marisaldi}, {Galli}, {Barbiellini}, {Longo},
  {Vallazza}, {Picozza}, {Morselli}, {Prest}, {Lipari}, {Zanello}, {Cattaneo},
  {Pittori}, {Verrecchia}, {Preger}, {Giommi}, \&
  {Salotti}}]{2008ATel.1574....1C}
{Costa}, E., {Del Monte}, E., {Donnarumma}, I., {et~al.} 2008, The Astronomer's
  Telegram, 1574

\bibitem[{{Cui}(2004)}]{2004ApJ...605..662C}
{Cui}, W. 2004, \apj, 605, 662

\bibitem[{{Djannati-Atai } {et~al.}(1999){Djannati-Atai }, {Piron}, {Barrau},
  {Iacoucci}, {Punch}, {Tavernet}, {Bazer-Bachi}, {Cabot}, {Chounet},
  {Debiais}, {Degrange}, {Dezalay}, {Dumora}, {Espigat}, {Fabre}, {Fleury},
  {Fontaine}, {Ghesqui{\`e}re}, {Goret}, {Gouiffes}, {Grenier}, {Le Bohec},
  {Malet}, {Meynadier}, {Mohanty}, {Nuss}, {Par{\'e}}, {Qu{\'e}bert}, {Ragan},
  {Renault}, {Rivoal}, {Rob}, {Schahmaneche}, \& {Smith}}]{1999A&A...350...17D}
{Djannati-Atai }, A., {Piron}, F., {Barrau}, A., {et~al.} 1999, \aap, 350, 17

\bibitem[{{Edelson} \& {Krolik}(1988)}]{1988ApJ...333..646E}
{Edelson}, R.~A., \& {Krolik}, J.~H. 1988, \apj, 333, 646

\bibitem[{{Elvis} {et~al.}(1992){Elvis}, {Plummer}, {Schachter}, \&
  {Fabbiano}}]{1992ApJS...80..257E}
{Elvis}, M., {Plummer}, D., {Schachter}, J., \& {Fabbiano}, G. 1992, \apjs, 80,
  257

\bibitem[{{Falomo} {et~al.}(1993){Falomo}, {Pesce}, \&
  {Treves}}]{1993ApJ...411L..63F}
{Falomo}, R., {Pesce}, J.~E., \& {Treves}, A. 1993, \apjl, 411, L63

\bibitem[{{Fan} \& {Lin}(1999)}]{1999ApJS..121..131F}
{Fan}, J.~H., \& {Lin}, R.~G. 1999, \apjs, 121, 131

\bibitem[{{Farina} {et~al.}(2016){Farina}, {Fumagalli}, {Decarli}, \&
  {Fanidakis}}]{2016MNRAS.455..618F}
{Farina}, E.~P., {Fumagalli}, M., {Decarli}, R., \& {Fanidakis}, N. 2016,
  \mnras, 455, 618

\bibitem[{{Fossati} {et~al.}(2008){Fossati}, {Buckley}, {Bond}, {Bradbury},
  {Carter-Lewis}, {Chow}, {Cui}, {Falcone}, {Finley}, {Gaidos}, {Grube},
  {Holder}, {Horan}, {Horns}, {Jordan}, {Kieda}, {Kildea}, {Krawczynski},
  {Krennrich}, {Lang}, {LeBohec}, {Lee}, {Moriarty}, {Ong}, {Petry}, {Quinn},
  {Sembroski}, {Wakely}, \& {Weekes}}]{2008ApJ...677..906F}
{Fossati}, G., {Buckley}, J.~H., {Bond}, I.~H., {et~al.} 2008, \apj, 677, 906

\bibitem[{{Fraija} {et~al.}(2015){Fraija}, {Cabrera}, {Ben{\'{\i}}tez}, \&
  {Hiriart}}]{2015arXiv150801438F}
{Fraija}, N., {Cabrera}, J.~I., {Ben{\'{\i}}tez}, E., \& {Hiriart}, D. 2015,
  ArXiv e-prints, arXiv:1508.01438

\bibitem[{{Furniss} {et~al.}(2015){Furniss}, {Noda}, {Boggs}, {Chiang},
  {Christensen}, {Craig}, {Giommi}, {Hailey}, {Harisson}, {Madejski}, \&
  et~al.}]{2015ApJ...812...65F}
{Furniss}, A., {Noda}, K., {Boggs}, S., {et~al.} 2015, \apj, 812, 65

\bibitem[{{Gaur} {et~al.}(2010){Gaur}, {Gupta}, {Lachowicz}, \&
  {Wiita}}]{2010ApJ...718..279G}
{Gaur}, H., {Gupta}, A.~C., {Lachowicz}, P., \& {Wiita}, P.~J. 2010, \apj, 718,
  279

\bibitem[{{Gaur} {et~al.}(2012{\natexlab{a}}){Gaur}, {Gupta}, \&
  {Wiita}}]{2012AJ....143...23G}
{Gaur}, H., {Gupta}, A.~C., \& {Wiita}, P.~J. 2012{\natexlab{a}}, \aj, 143, 23

\bibitem[{{Gaur} {et~al.}(2012{\natexlab{b}}){Gaur}, {Gupta}, {Strigachev},
  {Bachev}, {Semkov}, {Wiita}, {Peneva}, {Boeva}, {Slavcheva-Mihova}, {Mihov},
  {Latev}, \& {Pandey}}]{2012MNRAS.425.3002G}
{Gaur}, H., {Gupta}, A.~C., {Strigachev}, A., {et~al.} 2012{\natexlab{b}},
  \mnras, 425, 3002

\bibitem[{{Gaur} {et~al.}(2012{\natexlab{c}}){Gaur}, {Gupta}, {Strigachev},
  {Bachev}, {Semkov}, {Wiita}, {Peneva}, {Boeva}, {Kacharov}, {Mihov}, \&
  {Ovcharov}}]{2012MNRAS.420.3147G}
---. 2012{\natexlab{c}}, \mnras, 420, 3147

\bibitem[{{Gaur} {et~al.}(2015){Gaur}, {Gupta}, {Bachev}, {Strigachev},
  {Semkov}, {Wiita}, {Volvach}, {Gu}, {Agarwal}, {Agudo}, {Aller}, {Aller},
  {Kurtanidze}, {Kurtanidze}, {L{\"a}hteenm{\"a}ki}, {Peneva}, {Nikolashvili},
  {Sigua}, {Tornikoski}, \& {Volvach}}]{2015A&A...582A.103G}
{Gaur}, H., {Gupta}, A.~C., {Bachev}, R., {et~al.} 2015, \aap, 582, A103

\bibitem[{{Ghisellini} \& {Tavecchio}(2008)}]{2008MNRAS.386L..28G}
{Ghisellini}, G., \& {Tavecchio}, F. 2008, \mnras, 386, L28

\bibitem[{{Giannios} {et~al.}(2009){Giannios}, {Uzdensky}, \&
  {Begelman}}]{2009MNRAS.395L..29G}
{Giannios}, D., {Uzdensky}, D.~A., \& {Begelman}, M.~C. 2009, \mnras, 395, L29

\bibitem[{{Giebels} {et~al.}(2002){Giebels}, {Bloom}, {Focke}, {Godfrey},
  {Madejski}, {Reilly}, {Parkinson}, {Shabad}, {Bandyopadhyay}, {Fritz},
  {Hertz}, {Kowalski}, {Lovellette}, {Ray}, {Wolff}, {Wood}, {Yentis}, \&
  {Scargle}}]{2002ApJ...571..763G}
{Giebels}, B., {Bloom}, E.~D., {Focke}, W., {et~al.} 2002, \apj, 571, 763

\bibitem[{{Giroletti} {et~al.}(2004){Giroletti}, {Giovannini}, {Feretti},
  {Cotton}, {Edwards}, {Lara}, {Marscher}, {Mattox}, {Piner}, \&
  {Venturi}}]{2004ApJ...600..127G}
{Giroletti}, M., {Giovannini}, G., {Feretti}, L., {et~al.} 2004, \apj, 600, 127

\bibitem[{{Gopal-Krishna} \& {Wiita}(1992)}]{1992A&A...259..109G}
{Gopal-Krishna}, \& {Wiita}, P.~J. 1992, \aap, 259, 109

\bibitem[{{Gopal-Krishna} {et~al.}(2006){Gopal-Krishna}, {Wiita}, \&
  {Dhurde}}]{2006MNRAS.369.1287G}
{Gopal-Krishna}, {Wiita}, P.~J., \& {Dhurde}, S. 2006, \mnras, 369, 1287

\bibitem[{{Gupta}(2011)}]{2011JApA...32..155G}
{Gupta}, A.~C. 2011, Journal of Astrophysics and Astronomy, 32, 155

\bibitem[{{Gupta} {et~al.}(2008{\natexlab{a}}){Gupta}, {Acharya}, {Bose},
  {Chitnis}, \& {Fan}}]{2008ChJAA...8..395G}
{Gupta}, A.~C., {Acharya}, B.~S., {Bose}, D., {Chitnis}, V.~R., \& {Fan}, J.-H.
  2008{\natexlab{a}}, \cjaa, 8, 395

\bibitem[{{Gupta} {et~al.}(2004){Gupta}, {Banerjee}, {Ashok}, \&
  {Joshi}}]{2004A&A...422..505G}
{Gupta}, A.~C., {Banerjee}, D.~P.~K., {Ashok}, N.~M., \& {Joshi}, U.~C. 2004,
  \aap, 422, 505

\bibitem[{{Gupta} {et~al.}(2008{\natexlab{b}}){Gupta}, {Deng}, {Joshi}, {Bai},
  \& {Lee}}]{2008NewA...13..375G}
{Gupta}, A.~C., {Deng}, W.~G., {Joshi}, U.~C., {Bai}, J.~M., \& {Lee}, M.~G.
  2008{\natexlab{b}}, \na, 13, 375

\bibitem[{{Gupta} \& {Joshi}(2005)}]{2005A&A...440..855G}
{Gupta}, A.~C., \& {Joshi}, U.~C. 2005, \aap, 440, 855

\bibitem[{{Gupta} {et~al.}(2016{\natexlab{a}}){Gupta}, {Kalita}, {Gaur}, \&
  {Duorah}}]{2016MNRAS.462.1508G}
{Gupta}, A.~C., {Kalita}, N., {Gaur}, H., \& {Duorah}, K. 2016{\natexlab{a}},
  \mnras, 462, 1508

\bibitem[{{Gupta} {et~al.}(2016{\natexlab{b}}){Gupta}, {Agarwal}, {Bhagwan},
  {Strigachev}, {Bachev}, {Semkov}, {Gaur}, {Damljanovic}, {Vince}, \&
  {Wiita}}]{2016MNRAS.458.1127G}
{Gupta}, A.~C., {Agarwal}, A., {Bhagwan}, J., {et~al.} 2016{\natexlab{b}},
  \mnras, 458, 1127

\bibitem[{{Gupta} {et~al.}(2012){Gupta}, {Pandey}, {Singh}, {Rani}, {Pan},
  {Fan}, \& {Gupta}}]{2012NewA...17....8G}
{Gupta}, S.~P., {Pandey}, U.~S., {Singh}, K., {et~al.} 2012, \na, 17, 8

\bibitem[{{Harrison} {et~al.}(2013){Harrison}, {Craig}, {Christensen},
  {Hailey}, {Zhang}, {Boggs}, {Stern}, {Cook}, {Forster}, {Giommi},
  {Grefenstette}, {Kim}, {Kitaguchi}, {Koglin}, {Madsen}, {Mao}, {Miyasaka},
  {Mori}, {Perri}, {Pivovaroff}, {Puccetti}, {Rana}, {Westergaard}, {Willis},
  {Zoglauer}, {An}, {Bachetti}, {Barri{\`e}re}, {Bellm}, {Bhalerao},
  {Brejnholt}, {Fuerst}, {Liebe}, {Markwardt}, {Nynka}, {Vogel}, {Walton},
  {Wik}, {Alexander}, {Cominsky}, {Hornschemeier}, {Hornstrup}, {Kaspi},
  {Madejski}, {Matt}, {Molendi}, {Smith}, {Tomsick}, {Ajello}, {Ballantyne},
  {Balokovi{\'c}}, {Barret}, {Bauer}, {Blandford}, {Brandt}, {Brenneman},
  {Chiang}, {Chakrabarty}, {Chenevez}, {Comastri}, {Dufour}, {Elvis}, {Fabian},
  {Farrah}, {Fryer}, {Gotthelf}, {Grindlay}, {Helfand}, {Krivonos}, {Meier},
  {Miller}, {Natalucci}, {Ogle}, {Ofek}, {Ptak}, {Reynolds}, {Rigby},
  {Tagliaferri}, {Thorsett}, {Treister}, \& {Urry}}]{2013ApJ...770..103H}
{Harrison}, F.~A., {Craig}, W.~W., {Christensen}, F.~E., {et~al.} 2013, \apj,
  770, 103

\bibitem[{{Horan} {et~al.}(2009){Horan}, {Acciari}, {Bradbury}, {Buckley},
  {Bugaev}, {Byrum}, {Cannon}, {Celik}, {Cesarini}, {Chow}, {Ciupik}, {Cogan},
  {Falcone}, {Fegan}, {Finley}, {Fortin}, {Fortson}, {Gall}, {Gillanders},
  {Grube}, {Gyuk}, {Hanna}, {Hays}, {Kertzman}, {Kildea}, {Konopelko},
  {Krawczynski}, {Krennrich}, {Lang}, {Lee}, {Moriarty}, {Nagai}, {Niemiec},
  {Ong}, {Perkins}, {Pohl}, {Quinn}, {Reynolds}, {Rose}, {Sembroski}, {Smith},
  {Steele}, {Swordy}, {Toner}, {Vassiliev}, {Wakely}, {Weekes}, {White},
  {Williams}, {Wood}, {Zitzer}, {Aller}, {Aller}, {Baker}, {Barnaby}, {Carini},
  {Charlot}, {Dumm}, {Fields}, {Hovatta}, {Jordan}, {Kovalev}, {Kovalev},
  {Krimm}, {Kurtanidze}, {L{\"a}hteenm{\"a}ki}, {LeCampion}, {Maune},
  {Montaruli}, {Sadun}, {Smith}, {Tornikoski}, {Turunen}, \&
  {Walters}}]{2009ApJ...695..596H}
{Horan}, D., {Acciari}, V.~A., {Bradbury}, S.~M., {et~al.} 2009, \apj, 695, 596

\bibitem[{{Hovatta} {et~al.}(2015){Hovatta}, {Petropoulou}, {Richards},
  {Giannios}, {Wiik}, {Balokovi{\'c}}, {L{\"a}hteenm{\"a}ki}, {Lott},
  {Max-Moerbeck}, {Ramakrishnan}, \& {Readhead}}]{2015MNRAS.448.3121H}
{Hovatta}, T., {Petropoulou}, M., {Richards}, J.~L., {et~al.} 2015, \mnras,
  448, 3121

\bibitem[{{Isobe} {et~al.}(2015){Isobe}, {Sato}, {Ueda}, {Hayashida},
  {Shidatsu}, {Kawamuro}, {Ueno}, {Sugizaki}, {Sugimoto}, {Mihara}, {Matsuoka},
  \& {Negoro}}]{2015ApJ...798...27I}
{Isobe}, N., {Sato}, R., {Ueda}, Y., {et~al.} 2015, \apj, 798, 27

\bibitem[{{Kalita} {et~al.}(2015){Kalita}, {Gupta}, {Wiita}, {Bhagwan}, \&
  {Duorah}}]{2015MNRAS.451.1356K}
{Kalita}, N., {Gupta}, A.~C., {Wiita}, P.~J., {Bhagwan}, J., \& {Duorah}, K.
  2015, \mnras, 451, 1356

\bibitem[{{Kataoka} \& {Stawarz}(2016)}]{2016arXiv160603659K}
{Kataoka}, J., \& {Stawarz}, L. 2016, ArXiv e-prints, arXiv:1606.03659

\bibitem[{{Kataoka} {et~al.}(2000){Kataoka}, {Takahashi}, {Makino}, {Inoue},
  {Madejski}, {Tashiro}, {Urry}, \& {Kubo}}]{2000ApJ...528..243K}
{Kataoka}, J., {Takahashi}, T., {Makino}, F., {et~al.} 2000, \apj, 528, 243

\bibitem[{{Kerrick} {et~al.}(1995){Kerrick}, {Akerlof}, {Biller}, {Buckley},
  {Cawley}, {Chantell}, {Connaughton}, {Fegan}, {Fennell}, {Gaidos}, {Hillas},
  {Lamb}, {Lewis}, {Meyer}, {McEnery}, {Mohanty}, {Quinn}, {Rovero}, {Rose},
  {Schubnell}, {Sembroski}, {Urban}, {Watson}, {Weekes}, {West}, {Wilson}, \&
  {Zweerink}}]{1995ApJ...438L..59K}
{Kerrick}, A.~D., {Akerlof}, C.~W., {Biller}, S.~D., {et~al.} 1995, \apjl, 438,
  L59

\bibitem[{{Lachowicz} {et~al.}(2009){Lachowicz}, {Gupta}, {Gaur}, \&
  {Wiita}}]{2009A&A...506L..17L}
{Lachowicz}, P., {Gupta}, A.~C., {Gaur}, H., \& {Wiita}, P.~J. 2009, \aap, 506,
  L17

\bibitem[{{Li} {et~al.}(2016){Li}, {Jiang}, {Guo}, {Chen}, \&
  {Yi}}]{2016PASP..128g4101L}
{Li}, H.~Z., {Jiang}, Y.~G., {Guo}, D.~F., {Chen}, X., \& {Yi}, T.~F. 2016,
  \pasp, 128, 074101

\bibitem[{{Lico} {et~al.}(2012){Lico}, {Giroletti}, {Orienti}, {Giovannini},
  {Cotton}, {Edwards}, {Fuhrmann}, {Krichbaum}, {Sokolovsky}, {Kovalev},
  {Jorstad}, {Marscher}, {Kino}, {Paneque}, {Perez-Torres}, \&
  {Piner}}]{2012A&A...545A.117L}
{Lico}, R., {Giroletti}, M., {Orienti}, M., {et~al.} 2012, \aap, 545, A117

\bibitem[{{MAGIC Collaboration} {et~al.}(2016){MAGIC Collaboration}, {Ahnen},
  {Ansoldi}, {Antonelli}, {Antoranz}, {Babic}, {Banerjee}, {Bangale}, {Barres
  de Almeida}, {Barrio}, {Becerra Gonz{\'a}lez}, {Bednarek}, {Bernardini},
  {Biasuzzi}, {Biland}, {Blanch}, {Bonnefoy}, {Bonnoli}, {Borracci}, {Bretz},
  {Buson}, {Carosi}, {Chatterjee}, {Clavero}, {Colin}, {Colombo}, {Contreras},
  {Cortina}, {Covino}, {Da Vela}, {Dazzi}, {De Angelis}, {De Lotto}, {de
  O{\~n}a Wilhelmi}, {Di Pierro}, {Dom{\'{\i}}nguez}, {Dominis Prester},
  {Dorner}, {Doro}, {Einecke}, {Eisenacher Glawion}, {Elsaesser},
  {Fern{\'a}ndez-Barral}, {Fidalgo}, {Fonseca}, {Font}, {Frantzen}, {Fruck},
  {Galindo}, {Garc{\'{\i}}a L{\'o}pez}, {Garczarczyk}, {Garrido Terrats},
  {Gaug}, {Giammaria}, {Godinovi{\'c}}, {Gonz{\'a}lez Mu{\~n}oz}, {Gora},
  {Guberman}, {Hadasch}, {Hahn}, {Hanabata}, {Hayashida}, {Herrera}, {Hose},
  {Hrupec}, {Hughes}, {Idec}, {Kodani}, {Konno}, {Kubo}, {Kushida}, {La
  Barbera}, {Lelas}, {Lindfors}, {Lombardi}, {Longo}, {L{\'o}pez},
  {L{\'o}pez-Coto}, {Majumdar}, {Makariev}, {Mallot}, {Maneva}, {Manganaro},
  {Mannheim}, {Maraschi}, {Marcote}, {Mariotti}, {Mart{\'{\i}}nez}, {Mazin},
  {Menzel}, {Miranda}, {Mirzoyan}, {Moralejo}, {Moretti}, {Nakajima},
  {Neustroev}, {Niedzwiecki}, {Nievas Rosillo}, {Nilsson}, {Nishijima}, {Noda},
  {Nogu{\'e}s}, {Orito}, {Overkemping}, {Paiano}, {Palacio}, {Palatiello},
  {Paneque}, {Paoletti}, {Paredes}, {Paredes-Fortuny}, {Pedaletti}, {Perri},
  {Persic}, {Poutanen}, {Prada Moroni}, {Prandini}, {Puljak}, {Rhode},
  {Rib{\'o}}, {Rico}, {Rodriguez Garcia}, {Saito}, {Satalecka}, {Schultz},
  {Schweizer}, {Shore}, {Sillanp{\"a}{\"a}}, {Sitarek}, {Snidaric},
  {Sobczynska}, {Stamerra}, {Steinbring}, {Strzys}, {Takalo}, {Takami},
  {Tavecchio}, {Temnikov}, {Terzi{\'c}}, {Tescaro}, {Teshima}, {Thaele},
  {Torres}, {Toyama}, {Treves}, {Verguilov}, {Vovk}, {Ward}, {Will}, {Wu},
  {Zanin}, {Blinov}, {Chen}, {Efimova}, {Forn{\'e}}, {Grishina}, {Hovatta},
  {Jordan}, {Kimeridze}, {Kopatskaya}, {Koptelova}, {Kurtanidze}, {Kurtanidze},
  {L{\"a}hteenma{\"a}ki}, {Larionov}, {Larionova}, {Larionova}, {Ligustri},
  {Lin}, {McBreen}, {Morozova}, {Nikolashvili}, {Raiteri}, {Ros}, {Sadun},
  {Sigua}, {Tornikoski}, {Troitsky}, \& {Villata}}]{2016arXiv160509017M}
{MAGIC Collaboration}, {Ahnen}, M.~L., {Ansoldi}, S., {et~al.} 2016, ArXiv
  e-prints, arXiv:1605.09017

\bibitem[{{Mangalam} \& {Wiita}(1993)}]{1993ApJ...406..420M}
{Mangalam}, A.~V., \& {Wiita}, P.~J. 1993, \apj, 406, 420

\bibitem[{{Marcha} {et~al.}(1996){Marcha}, {Browne}, {Impey}, \&
  {Smith}}]{1996MNRAS.281..425M}
{Marcha}, M.~J.~M., {Browne}, I.~W.~A., {Impey}, C.~D., \& {Smith}, P.~S. 1996,
  \mnras, 281, 425

\bibitem[{{Marscher}(2014)}]{2014ApJ...780...87M}
{Marscher}, A.~P. 2014, \apj, 780, 87

\bibitem[{{Marscher} \& {Gear}(1985)}]{1985ApJ...298..114M}
{Marscher}, A.~P., \& {Gear}, W.~K. 1985, \apj, 298, 114

\bibitem[{{Massaro} {et~al.}(2008){Massaro}, {Tramacere}, {Cavaliere}, {Perri},
  \& {Giommi}}]{2008A&A...478..395M}
{Massaro}, F., {Tramacere}, A., {Cavaliere}, A., {Perri}, M., \& {Giommi}, P.
  2008, \aap, 478, 395

\bibitem[{{Neronov} {et~al.}(2012){Neronov}, {Semikoz}, \&
  {Taylor}}]{2012A&A...541A..31N}
{Neronov}, A., {Semikoz}, D., \& {Taylor}, A.~M. 2012, \aap, 541, A31

\bibitem[{{Paliya} {et~al.}(2015){Paliya}, {B{\"o}ttcher}, {Diltz}, {Stalin},
  {Sahayanathan}, \& {Ravikumar}}]{2015ApJ...811..143P}
{Paliya}, V.~S., {B{\"o}ttcher}, M., {Diltz}, C., {et~al.} 2015, \apj, 811, 143

\bibitem[{{Perlman} {et~al.}(1996){Perlman}, {Stocke}, {Schachter}, {Elvis},
  {Ellingson}, {Urry}, {Potter}, {Impey}, \&
  {Kolchinsky}}]{1996ApJS..104..251P}
{Perlman}, E.~S., {Stocke}, J.~T., {Schachter}, J.~F., {et~al.} 1996, \apjs,
  104, 251

\bibitem[{{Pian} {et~al.}(1998){Pian}, {Vacanti}, {Tagliaferri}, {Ghisellini},
  {Maraschi}, {Treves}, {Urry}, {Fiore}, {Giommi}, {Palazzi}, {Chiappetti}, \&
  {Sambruna}}]{1998ApJ...492L..17P}
{Pian}, E., {Vacanti}, G., {Tagliaferri}, G., {et~al.} 1998, \apjl, 492, L17

\bibitem[{{Piner} \& {Edwards}(2004)}]{2004ApJ...600..115P}
{Piner}, B.~G., \& {Edwards}, P.~G. 2004, \apj, 600, 115

\bibitem[{{Pittori} {et~al.}(2008){Pittori}, {Cutini}, {Gasparrini},
  {Verrecchia}, {Colafrancesco}, {Giommi}, {Vercellone}, {Chen}, {Giuliani},
  {Donnarumma}, {D'Ammando}, {Pacciani}, {Pucella}, {Vittorini}, {Longo},
  {Bulgarelli}, {Mereghetti}, {Pellizzoni}, {Perotti}, {Fornari}, {Fiorini},
  {Caraveo}, {Zambra}, {Gianotti}, {Trifoglio}, {Di Cocco}, {Labanti},
  {Fuschino}, {Marisaldi}, {Galli}, {Tavani}, {Costa}, {Feroci}, {Del Monte},
  {Lazzarotto}, {Piano}, {Soffitta}, {Evangelista}, {Lapshov}, {Rapisarda},
  {Argan}, {Trois}, {de Paris}, {Barbiellini}, {Vallazza}, {Picozza},
  {Morselli}, {Prest}, {Lipari}, {Zanello}, {Cattaneo}, {Salotti}, \&
  {Grube}}]{2008ATel.1583....1P}
{Pittori}, C., {Cutini}, S., {Gasparrini}, D., {et~al.} 2008, The Astronomer's
  Telegram, 1583

\bibitem[{{Pollack} {et~al.}(2016){Pollack}, {Pauls}, \&
  {Wiita}}]{2016ApJ...820...12P}
{Pollack}, M., {Pauls}, D., \& {Wiita}, P.~J. 2016, \apj, 820, 12

\bibitem[{{Punch} {et~al.}(1992){Punch}, {Akerlof}, {Cawley}, {Chantell},
  {Fegan}, {Fennell}, {Gaidos}, {Hagan}, {Hillas}, {Jiang}, {Kerrick}, {Lamb},
  {Lawrence}, {Lewis}, {Meyer}, {Mohanty}, {O'Flaherty}, {Reynolds}, {Rovero},
  {Schubnell}, {Sembroski}, {Weekes}, \& {Wilson}}]{1992Natur.358..477P}
{Punch}, M., {Akerlof}, C.~W., {Cawley}, M.~F., {et~al.} 1992, \nat, 358, 477

\bibitem[{{Quinn} {et~al.}(1996){Quinn}, {Akerlof}, {Biller}, {Buckley},
  {Carter-Lewis}, {Cawley}, {Catanese}, {Connaughton}, {Fegan}, {Finley},
  {Gaidos}, {Hillas}, {Lamb}, {Krennrich}, {Lessard}, {McEnery}, {Meyer},
  {Mohanty}, {Rodgers}, {Rose}, {Sembroski}, {Schubnell}, {Weekes}, {Wilson},
  \& {Zweerink}}]{1996ApJ...456L..83Q}
{Quinn}, J., {Akerlof}, C.~W., {Biller}, S., {et~al.} 1996, \apjl, 456, L83

\bibitem[{{Racero} \& {de la Calle}(2013)}]{2013EPJWC..6104020R}
{Racero}, E., \& {de la Calle}, I. 2013, in European Physical Journal Web of
  Conferences, Vol.~61, European Physical Journal Web of Conferences, 04020

\bibitem[{{Rani} {et~al.}(2011){Rani}, {Gupta}, {Joshi}, {Ganesh}, \&
  {Wiita}}]{2011MNRAS.413.2157R}
{Rani}, B., {Gupta}, A.~C., {Joshi}, U.~C., {Ganesh}, S., \& {Wiita}, P.~J.
  2011, \mnras, 413, 2157

\bibitem[{{Ravasio} {et~al.}(2004){Ravasio}, {Tagliaferri}, {Ghisellini}, \&
  {Tavecchio}}]{2004A&A...424..841R}
{Ravasio}, M., {Tagliaferri}, G., {Ghisellini}, G., \& {Tavecchio}, F. 2004,
  \aap, 424, 841

\bibitem[{{Samuelson} {et~al.}(1998){Samuelson}, {Biller}, {Bond}, {Boyle},
  {Bradbury}, {Breslin}, {Buckley}, {Burdett}, {Gordo}, {Carter-Lewis},
  {Catanese}, {Cawley}, {Fegan}, {Finley}, {Gaidos}, {Hall}, {Hillas},
  {Krennrich}, {Lamb}, {Lessard}, {McEnery}, {Masterson}, {Quinn}, {Rodgers},
  {Rose}, {Sembroski}, {Srinivasan}, {Vassiliev}, {Weekes}, \&
  {Zweerink}}]{1998ApJ...501L..17S}
{Samuelson}, F.~W., {Biller}, S.~D., {Bond}, I.~H., {et~al.} 1998, \apjl, 501,
  L17

\bibitem[{{Sandrinelli} {et~al.}(2016){Sandrinelli}, {Covino}, {Dotti}, \&
  {Treves}}]{2016AJ....151...54S}
{Sandrinelli}, A., {Covino}, S., {Dotti}, M., \& {Treves}, A. 2016, \aj, 151,
  54

\bibitem[{{Sinha} {et~al.}(2015){Sinha}, {Shukla}, {Misra}, {Chitnis}, {Rao},
  \& {Acharya}}]{2015A&A...580A.100S}
{Sinha}, A., {Shukla}, A., {Misra}, R., {et~al.} 2015, \aap, 580, A100

\bibitem[{{Sinha} {et~al.}(2016){Sinha}, {Shukla}, {Saha}, {Acharya},
  {Anupama}, {Bhattacharjee}, {Britto}, {Chitnis}, {Prabhu}, {Singh}, \&
  {Vishwanath}}]{2016A&A...591A..83S}
{Sinha}, A., {Shukla}, A., {Saha}, L., {et~al.} 2016, \aap, 591, A83

\bibitem[{{Smith} {et~al.}(2008){Smith}, {Abdo}, {Allen}, {Berley}, {Casanova},
  {Chen}, {Coyne}, {Dingus}, {Ellsworth}, {Fleysher}, {Fleysher}, {Gonzalez},
  {Goodman}, {Hays}, {Hoffman}, {Hopper}, {H{\"u}ntemeyer}, {Kolterman},
  {Lansdell}, {Linnemann}, {McEnery}, {Mincer}, {Nemethy}, {Noyes}, {Ryan},
  {Saz Parkinson}, {Shoup}, {Sinnis}, {Sullivan}, {Vasileiou}, {Walker},
  {Williams}, {Zu}, \& {Yodh}}]{2008ICRC....3..973S}
{Smith}, A.~J., {Abdo}, A.~A., {Allen}, B., {et~al.} 2008, International Cosmic
  Ray Conference, 3, 973

\bibitem[{{Stein} {et~al.}(1976){Stein}, {Odell}, \&
  {Strittmatter}}]{1976ARA&A..14..173S}
{Stein}, W.~A., {Odell}, S.~L., \& {Strittmatter}, P.~A. 1976, \araa, 14, 173

\bibitem[{{Stocke} {et~al.}(1991){Stocke}, {Morris}, {Gioia}, {Maccacaro},
  {Schild}, {Wolter}, {Fleming}, \& {Henry}}]{1991ApJS...76..813S}
{Stocke}, J.~T., {Morris}, S.~L., {Gioia}, I.~M., {et~al.} 1991, \apjs, 76, 813

\bibitem[{{Tagliaferri} {et~al.}(2003){Tagliaferri}, {Ravasio}, {Ghisellini},
  {Tavecchio}, {Giommi}, {Massaro}, {Nesci}, \& {Tosti}}]{2003A&A...412..711T}
{Tagliaferri}, G., {Ravasio}, M., {Ghisellini}, G., {et~al.} 2003, \aap, 412,
  711

\bibitem[{{Takahashi} {et~al.}(1994){Takahashi}, {Tashiro}, {Kamae},
  {Makishima}, {Kii}, {Makino}, {Ohashi}, {Takahara}, {Yamasaki}, {Hartman},
  {Madejski}, {Lamb}, \& {Weekes}}]{1994IAUC.5993....2T}
{Takahashi}, T., {Tashiro}, M., {Kamae}, T., {et~al.} 1994, \iaucirc, 5993

\bibitem[{{Takahashi} {et~al.}(1995){Takahashi}, {Makino}, {Kii}, {Tashiro},
  {Kubo}, {Makishima}, {Ohashi}, {Yamasaki}, {Madejski}, {Buckley}, {Quinn},
  {Weekes}, {Catanese}, {Carter-Lewis}, {Krennrich}, {Lamb}, {Zweerink},
  {Akerlof}, {Meyer}, {Schubnell}, {Gaidos}, {Sembroski}, {Wilson}, {Fegan},
  {Lessard}, {McEnery}, {Cawley}, {Biller}, {Hillas}, {Rodgers}, \&
  {Rose}}]{1995IAUC.6167....1T}
{Takahashi}, T., {Makino}, F., {Kii}, T., {et~al.} 1995, \iaucirc, 6167

\bibitem[{{Ter{\"a}sranta} {et~al.}(2005){Ter{\"a}sranta}, {Wiren}, {Koivisto},
  {Saarinen}, \& {Hovatta}}]{2005A&A...440..409T}
{Ter{\"a}sranta}, H., {Wiren}, S., {Koivisto}, P., {Saarinen}, V., \&
  {Hovatta}, T. 2005, \aap, 440, 409

\bibitem[{{Ter{\"a}sranta} {et~al.}(2004){Ter{\"a}sranta}, {Achren}, {Hanski},
  {Heikkil{\"a}}, {Holopainen}, {Joutsamo}, {Juhola}, {Karlamaa}, {Katajainen},
  {Kein{\"a}nen}, {Koivisto}, {Koskimies}, {K{\"o}n{\"o}nen}, {Lainela},
  {L{\"a}htenm{\"a}ki}, {M{\"a}kinen}, {Niemel{\"a}}, {Nurmi}, {Pursimo},
  {Rekola}, {Savolainen}, {Tornikoski}, {Torppa}, {Valtonen}, {Varjonen},
  {Vilenius}, {Virtanen}, \& {Wiren}}]{2004A&A...427..769T}
{Ter{\"a}sranta}, H., {Achren}, J., {Hanski}, M., {et~al.} 2004, \aap, 427, 769

\bibitem[{{Tramacere} {et~al.}(2009){Tramacere}, {Giommi}, {Perri},
  {Verrecchia}, \& {Tosti}}]{2009A&A...501..879T}
{Tramacere}, A., {Giommi}, P., {Perri}, M., {Verrecchia}, F., \& {Tosti}, G.
  2009, \aap, 501, 879

\bibitem[{{Urry} \& {Padovani}(1995)}]{1995PASP..107..803U}
{Urry}, C.~M., \& {Padovani}, P. 1995, \pasp, 107, 803

\bibitem[{{Urry} {et~al.}(1993){Urry}, {Maraschi}, {Edelson}, {Koratkar},
  {Krolik}, {Madejski}, {Pian}, {Pike}, {Reichert}, {Treves}, {Wamsteker},
  {Bohlin}, {Bregman}, {Brinkmann}, {Chiappetti}, {Courvoisier}, {Filippenko},
  {Fink}, {George}, {Kondo}, {Martin}, {Miller}, {O'Brien}, {Shull}, {Sitko},
  {Szymkowiak}, {Tagliaferri}, {Wagner}, \& {Warwick}}]{1993ApJ...411..614U}
{Urry}, C.~M., {Maraschi}, L., {Edelson}, R., {et~al.} 1993, \apj, 411, 614

\bibitem[{{Ushio} {et~al.}(2009){Ushio}, {Tanaka}, {Madejski}, {Takahashi},
  {Hayashida}, {Kataoka}, {Mazin}, {R{\"u}gamer}, {Sato}, {Teshima}, {Wagner},
  \& {Yaji}}]{2009ApJ...699.1964U}
{Ushio}, M., {Tanaka}, T., {Madejski}, G., {et~al.} 2009, \apj, 699, 1964

\bibitem[{{Vaughan} {et~al.}(2003){Vaughan}, {Edelson}, {Warwick}, \&
  {Uttley}}]{2003MNRAS.345.1271V}
{Vaughan}, S., {Edelson}, R., {Warwick}, R.~S., \& {Uttley}, P. 2003, \mnras,
  345, 1271

\bibitem[{{Wagner} \& {Witzel}(1995)}]{1995ARA&A..33..163W}
{Wagner}, S.~J., \& {Witzel}, A. 1995, \araa, 33, 163

\bibitem[{{Woo} {et~al.}(2005){Woo}, {Urry}, {van der Marel}, {Lira}, \&
  {Maza}}]{2005ApJ...631..762W}
{Woo}, J.-H., {Urry}, C.~M., {van der Marel}, R.~P., {Lira}, P., \& {Maza}, J.
  2005, \apj, 631, 762

\bibitem[{{Xiong} {et~al.}(2016){Xiong}, {Zhang}, {Zhang}, {Yi}, {Bai}, {Wang},
  {Liu}, \& {Zheng}}]{2016ApJS..222...24X}
{Xiong}, D., {Zhang}, H., {Zhang}, X., {et~al.} 2016, \apjs, 222, 24

\bibitem[{{Zhang} {et~al.}(2014){Zhang}, {Zhao}, {Wang}, \&
  {Dai}}]{2014RAA....14..933Z}
{Zhang}, B.-K., {Zhao}, X.-Y., {Wang}, C.-X., \& {Dai}, B.-Z. 2014, Research in
  Astronomy and Astrophysics, 14, 933

\bibitem[{{Zhang} {et~al.}(2002){Zhang}, {Treves}, {Celotti}, {Chiappetti},
  {Fossati}, {Ghisellini}, {Maraschi}, {Pian}, {Tagliaferri}, \&
  {Tavecchio}}]{2002ApJ...572..762Z}
{Zhang}, Y.~H., {Treves}, A., {Celotti}, A., {et~al.} 2002, \apj, 572, 762

\end{thebibliography}

\end{document}